\pdfoutput=1
\documentclass[aps,prc,reprint,superscriptaddress]{revtex4-1}

\usepackage{slashed,amsmath,amssymb,hyperref,textcomp,graphicx}
\hypersetup{colorlinks,%
            citecolor=black,%
            linkcolor=black,%
            filecolor=black,%
            urlcolor=blue,%
            pdfauthor={NEXT Collaboration},%
            pdftitle={Ionization and scintillation of nuclear recoils in gaseous xenon}}

\begin{document}

\title{Ionization and scintillation of nuclear recoils in gaseous xenon}

\author{J.~Renner}\email[]{jrenner@lbl.gov}\affiliation{Lawrence Berkeley National Laboratory (LBNL),1 Cyclotron Road, Berkeley, California 94720, USA}\affiliation{Department of Physics, University of California, Berkeley, CA 94720, USA}
\author{V.\,M.~Gehman}\affiliation{Lawrence Berkeley National Laboratory (LBNL),1 Cyclotron Road, Berkeley, California 94720, USA}
\author{A.~Goldschmidt}\affiliation{Lawrence Berkeley National Laboratory (LBNL),1 Cyclotron Road, Berkeley, California 94720, USA}
\author{H.\,S.~Matis}\affiliation{Lawrence Berkeley National Laboratory (LBNL),1 Cyclotron Road, Berkeley, California 94720, USA}
\author{T.~Miller}\affiliation{Lawrence Berkeley National Laboratory (LBNL),1 Cyclotron Road, Berkeley, California 94720, USA}
\author{Y.~Nakajima}\affiliation{Lawrence Berkeley National Laboratory (LBNL),1 Cyclotron Road, Berkeley, California 94720, USA}
\author{D.~Nygren}\affiliation{Lawrence Berkeley National Laboratory (LBNL),1 Cyclotron Road, Berkeley, California 94720, USA}
\author{C.\,A.\,B.~Oliveira}\affiliation{Lawrence Berkeley National Laboratory (LBNL),1 Cyclotron Road, Berkeley, California 94720, USA}
\author{D.~Shuman}\affiliation{Lawrence Berkeley National Laboratory (LBNL),1 Cyclotron Road, Berkeley, California 94720, USA}
\author{V.~\'Alvarez}\affiliation{Instituto de F\'isica Corpuscular (IFIC), CSIC \& Universitat de Val\`encia, Calle Catedr\'atico Jos\'e Beltr\'an, 2, 46980 Paterna, Valencia, Spain}
\author{F.\,I.\,G.~Borges}\affiliation{Departamento de Fisica, Universidade de Coimbra, Rua Larga, 3004-516 Coimbra, Portugal}
\author{S.~C\'arcel}\affiliation{Instituto de F\'isica Corpuscular (IFIC), CSIC \& Universitat de Val\`encia, Calle Catedr\'atico Jos\'e Beltr\'an, 2, 46980 Paterna, Valencia, Spain}
\author{J.~Castel}\affiliation{Laboratorio de F\'isica Nuclear y Astropart\'iculas, Universidad de Zaragoza, Calle Pedro Cerbuna 12, 50009 Zaragoza, Spain}
\author{S.~Cebri\'an}\affiliation{Laboratorio de F\'isica Nuclear y Astropart\'iculas, Universidad de Zaragoza, Calle Pedro Cerbuna 12, 50009 Zaragoza, Spain}
\author{A.~Cervera}\affiliation{Instituto de F\'isica Corpuscular (IFIC), CSIC \& Universitat de Val\`encia, Calle Catedr\'atico Jos\'e Beltr\'an, 2, 46980 Paterna, Valencia, Spain}
\author{C.\,A.\,N.~Conde}\affiliation{Departamento de Fisica, Universidade de Coimbra, Rua Larga, 3004-516 Coimbra, Portugal}
\author{T.~Dafni}\affiliation{Laboratorio de F\'isica Nuclear y Astropart\'iculas, Universidad de Zaragoza, Calle Pedro Cerbuna 12, 50009 Zaragoza, Spain}
\author{T.\,H.\,V.\,T.~Dias}\affiliation{Departamento de Fisica, Universidade de Coimbra, Rua Larga, 3004-516 Coimbra, Portugal}
\author{J.~D\'iaz}\affiliation{Instituto de F\'isica Corpuscular (IFIC), CSIC \& Universitat de Val\`encia, Calle Catedr\'atico Jos\'e Beltr\'an, 2, 46980 Paterna, Valencia, Spain}
\author{R.~Esteve}\affiliation{Instituto de Instrumentaci\'on para Imagen Molecular (I3M), Universitat Polit\`ecnica de Val\`encia, Camino de Vera, s/n, Edificio 8B, 46022 Valencia, Spain}
\author{P.~Evtoukhovitch}\affiliation{Joint Institute for Nuclear Research (JINR), Joliot-Curie 6, 141980 Dubna, Russia}
\author{L.\,M.\,P.~Fernandes}\affiliation{Departamento de Fisica, Universidade de Coimbra, Rua Larga, 3004-516 Coimbra, Portugal}
\author{P.~Ferrario}\affiliation{Instituto de F\'isica Corpuscular (IFIC), CSIC \& Universitat de Val\`encia, Calle Catedr\'atico Jos\'e Beltr\'an, 2, 46980 Paterna, Valencia, Spain}
\author{A.\,L.~Ferreira}\affiliation{Instituto de Instrumentaci\'on para Imagen Molecular (I3M), Universitat Polit\`ecnica de Val\`encia, Camino de Vera, s/n, Edificio 8B, 46022 Valencia, Spain}
\author{E.\,D.\,C.~Freitas}\affiliation{Departamento de Fisica, Universidade de Coimbra, Rua Larga, 3004-516 Coimbra, Portugal}
\author{A.~Gil}\affiliation{Instituto de F\'isica Corpuscular (IFIC), CSIC \& Universitat de Val\`encia, Calle Catedr\'atico Jos\'e Beltr\'an, 2, 46980 Paterna, Valencia, Spain}
\author{H.~G\'omez}\affiliation{Laboratorio de F\'isica Nuclear y Astropart\'iculas, Universidad de Zaragoza, Calle Pedro Cerbuna 12, 50009 Zaragoza, Spain}
\author{J.\,J.~G\'omez-Cadenas}\thanks{spokesperson}\affiliation{Instituto de F\'isica Corpuscular (IFIC), CSIC \& Universitat de Val\`encia, Calle Catedr\'atico Jos\'e Beltr\'an, 2, 46980 Paterna, Valencia, Spain}
\author{D.~Gonz\'alez-D\'iaz}\affiliation{Laboratorio de F\'isica Nuclear y Astropart\'iculas, Universidad de Zaragoza, Calle Pedro Cerbuna 12, 50009 Zaragoza, Spain}
\author{R.\,M.~Guti\'errez}\affiliation{Centro de Investigaciones, Universidad Antonio Nari\~{n}o, Carretera 3 Este No. 47A-15, Bogot\'{a}, Colombia}
\author{J.~Hauptman}\affiliation{Department of Physics and Astronomy, Iowa State University, 12 Physics Hall, Ames, Iowa 50011-3160, USA}
\author{J.\,A.~Hernando Morata}\affiliation{Instituto Gallego de F\'isica de Altas Energ\'ias (IGFAE), Univ.\ de Santiago de Compostela, Campus sur, R\'ua Xos\'e Mar\'ia Su\'arez N\'u\~nez, s/n, 15782 Santiago de Compostela, Spain}
\author{D.\,C.~Herrera}\affiliation{Laboratorio de F\'isica Nuclear y Astropart\'iculas, Universidad de Zaragoza, Calle Pedro Cerbuna 12, 50009 Zaragoza, Spain}
\author{F.\,J.~Iguaz}\affiliation{Laboratorio de F\'isica Nuclear y Astropart\'iculas, Universidad de Zaragoza, Calle Pedro Cerbuna 12, 50009 Zaragoza, Spain}
\author{I.\,G.~Irastorza}\affiliation{Laboratorio de F\'isica Nuclear y Astropart\'iculas, Universidad de Zaragoza, Calle Pedro Cerbuna 12, 50009 Zaragoza, Spain}
\author{M.\,A.~Jinete}\affiliation{Centro de Investigaciones, Universidad Antonio Nari\~{n}o, Carretera 3 Este No. 47A-15, Bogot\'{a}, Colombia}
\author{L.~Labarga}\affiliation{Departamento de F\'isica Te\'orica, Universidad Aut\'onoma de Madrid, Campus de Cantoblanco, 28049 Madrid, Spain}
\author{A.~Laing}\affiliation{Instituto de F\'isica Corpuscular (IFIC), CSIC \& Universitat de Val\`encia, Calle Catedr\'atico Jos\'e Beltr\'an, 2, 46980 Paterna, Valencia, Spain}
\author{I.~Liubarsky}\affiliation{Instituto de F\'isica Corpuscular (IFIC), CSIC \& Universitat de Val\`encia, Calle Catedr\'atico Jos\'e Beltr\'an, 2, 46980 Paterna, Valencia, Spain}
\author{J.\,A.\,M.~Lopes}\affiliation{Departamento de Fisica, Universidade de Coimbra, Rua Larga, 3004-516 Coimbra, Portugal}
\author{D.~Lorca}\affiliation{Instituto de F\'isica Corpuscular (IFIC), CSIC \& Universitat de Val\`encia, Calle Catedr\'atico Jos\'e Beltr\'an, 2, 46980 Paterna, Valencia, Spain}
\author{M.~Losada}\affiliation{Centro de Investigaciones, Universidad Antonio Nari\~{n}o, Carretera 3 Este No. 47A-15, Bogot\'{a}, Colombia}
\author{G.~Luz\'on}\affiliation{Laboratorio de F\'isica Nuclear y Astropart\'iculas, Universidad de Zaragoza, Calle Pedro Cerbuna 12, 50009 Zaragoza, Spain}
\author{A.~Mar\'i}\affiliation{Instituto de Instrumentaci\'on para Imagen Molecular (I3M), Universitat Polit\`ecnica de Val\`encia, Camino de Vera, s/n, Edificio 8B, 46022 Valencia, Spain}
\author{J.~Mart\'in-Albo}\affiliation{Instituto de F\'isica Corpuscular (IFIC), CSIC \& Universitat de Val\`encia, Calle Catedr\'atico Jos\'e Beltr\'an, 2, 46980 Paterna, Valencia, Spain}
\author{A.~Mart\'inez}\affiliation{Instituto de F\'isica Corpuscular (IFIC), CSIC \& Universitat de Val\`encia, Calle Catedr\'atico Jos\'e Beltr\'an, 2, 46980 Paterna, Valencia, Spain}
\author{A.~Moiseenko}\affiliation{Joint Institute for Nuclear Research (JINR), Joliot-Curie 6, 141980 Dubna, Russia}
\author{F.~Monrabal}\affiliation{Instituto de F\'isica Corpuscular (IFIC), CSIC \& Universitat de Val\`encia, Calle Catedr\'atico Jos\'e Beltr\'an, 2, 46980 Paterna, Valencia, Spain}
\author{M.~Monserrate}\affiliation{Instituto de F\'isica Corpuscular (IFIC), CSIC \& Universitat de Val\`encia, Calle Catedr\'atico Jos\'e Beltr\'an, 2, 46980 Paterna, Valencia, Spain}
\author{C.\,M.\,B.~Monteiro}\affiliation{Departamento de Fisica, Universidade de Coimbra, Rua Larga, 3004-516 Coimbra, Portugal}
\author{F.\,J.~Mora}\affiliation{Instituto de Instrumentaci\'on para Imagen Molecular (I3M), Universitat Polit\`ecnica de Val\`encia, Camino de Vera, s/n, Edificio 8B, 46022 Valencia, Spain}
\author{L.\,M.~Moutinho}\affiliation{Institute of Nanostructures, Nanomodelling and Nanofabrication (i3N), Universidade de Aveiro, Campus de Santiago, 3810-193 Aveiro, Portugal}
\author{J.~Mu\~noz Vidal}\affiliation{Instituto de F\'isica Corpuscular (IFIC), CSIC \& Universitat de Val\`encia, Calle Catedr\'atico Jos\'e Beltr\'an, 2, 46980 Paterna, Valencia, Spain}
\author{H.~Natal da Luz}\affiliation{Departamento de Fisica, Universidade de Coimbra, Rua Larga, 3004-516 Coimbra, Portugal}
\author{G.~Navarro}\affiliation{Centro de Investigaciones, Universidad Antonio Nari\~{n}o, Carretera 3 Este No. 47A-15, Bogot\'{a}, Colombia}
\author{M.~Nebot-Guinot}\affiliation{Instituto de F\'isica Corpuscular (IFIC), CSIC \& Universitat de Val\`encia, Calle Catedr\'atico Jos\'e Beltr\'an, 2, 46980 Paterna, Valencia, Spain}
\author{R.~Palma}\affiliation{Departamento de Mec\'anica de Medios Continuos y Teor\'ia de Estructuras, Univ.\ Polit\`ecnica de Val\`encia, Camino de Vera, s/n, 46071 Valencia, Spain}
\author{J.~P\'erez}\affiliation{Instituto de F\'isica Te\'orica (IFT), UAM/CSIC, Campus de Cantoblanco, 28049 Madrid, Spain}
\author{J.\,L.~P\'erez Aparicio}\affiliation{Dpto.\ de Mec\'anica de Medios Continuos y Teor\'ia de Estructuras, Univ.\ Polit\`ecnica de Val\`encia, Camino de Vera, s/n, 46071 Valencia, Spain}
\author{L.~Ripoll}\affiliation{Escola Polit\`ecnica Superior, Universitat de Girona, Av.~Montilivi, s/n, 17071 Girona, Spain}
\author{A.~Rodr\'iguez}\affiliation{Laboratorio de F\'isica Nuclear y Astropart\'iculas, Universidad de Zaragoza, Calle Pedro Cerbuna 12, 50009 Zaragoza, Spain}
\author{J.~Rodr\'iguez}\affiliation{Instituto de F\'isica Corpuscular (IFIC), CSIC \& Universitat de Val\`encia, Calle Catedr\'atico Jos\'e Beltr\'an, 2, 46980 Paterna, Valencia, Spain}
\author{F.\,P.~Santos}\affiliation{Departamento de Fisica, Universidade de Coimbra, Rua Larga, 3004-516 Coimbra, Portugal}
\author{J.\,M.\,F.~dos Santos}\affiliation{Departamento de Fisica, Universidade de Coimbra, Rua Larga, 3004-516 Coimbra, Portugal}
\author{L.~Segu\'{i}}\affiliation{Laboratorio de F\'isica Nuclear y Astropart\'iculas, Universidad de Zaragoza, Calle Pedro Cerbuna 12, 50009 Zaragoza, Spain}
\author{L.~Serra}\affiliation{Instituto de F\'isica Corpuscular (IFIC), CSIC \& Universitat de Val\`encia, Calle Catedr\'atico Jos\'e Beltr\'an, 2, 46980 Paterna, Valencia, Spain}
\author{A.~Sim\'on}\affiliation{Instituto de F\'isica Corpuscular (IFIC), CSIC \& Universitat de Val\`encia, Calle Catedr\'atico Jos\'e Beltr\'an, 2, 46980 Paterna, Valencia, Spain}
\author{C.~Sofka}\affiliation{Department of Physics and Astronomy, Texas A\&M University, College Station, Texas 77843-4242, USA}
\author{M.~Sorel}\affiliation{Instituto de F\'isica Corpuscular (IFIC), CSIC \& Universitat de Val\`encia, Calle Catedr\'atico Jos\'e Beltr\'an, 2, 46980 Paterna, Valencia, Spain}
\author{J.\,F.~Toledo}\affiliation{Instituto de Instrumentaci\'on para Imagen Molecular (I3M), Universitat Polit\`ecnica de Val\`encia, Camino de Vera, s/n, Edificio 8B, 46022 Valencia, Spain}
\author{A.~Tom\'as}\affiliation{Laboratorio de F\'isica Nuclear y Astropart\'iculas, Universidad de Zaragoza, Calle Pedro Cerbuna 12, 50009 Zaragoza, Spain}
\author{J.~Torrent}\affiliation{Escola Polit\`ecnica Superior, Universitat de Girona, Av.~Montilivi, s/n, 17071 Girona, Spain}
\author{Z.~Tsamalaidze}\affiliation{Joint Institute for Nuclear Research (JINR), Joliot-Curie 6, 141980 Dubna, Russia}
\author{J.\,F.\,C.\,A.~Veloso}\affiliation{Institute of Nanostructures, Nanomodelling and Nanofabrication (i3N), Universidade de Aveiro, Campus de Santiago, 3810-193 Aveiro, Portugal}
\author{J.\,A.~Villar}\affiliation{Laboratorio de F\'isica Nuclear y Astropart\'iculas, Universidad de Zaragoza, Calle Pedro Cerbuna 12, 50009 Zaragoza, Spain}
\author{R.\,C.~Webb}\affiliation{Department of Physics and Astronomy, Texas A\&M University, College Station, Texas 77843-4242, USA}
\author{J.~White}\thanks{deceased}\affiliation{Department of Physics and Astronomy, Texas A\&M University, College Station, Texas 77843-4242, USA}
\author{N.~Yahlali}\affiliation{Instituto de F\'isica Corpuscular (IFIC), CSIC \& Universitat de Val\`encia, Calle Catedr\'atico Jos\'e Beltr\'an, 2, 46980 Paterna, Valencia, Spain}
\collaboration{NEXT Collaboration}

\date{\today}

\begin{abstract}
Ionization and scintillation produced by nuclear recoils in gaseous xenon at approximately 14 bar have been simultaneously observed in an electroluminescent
time projection chamber.  Neutrons from radioisotope $\alpha$-Be neutron sources were used to induce xenon nuclear recoils, and the observed recoil spectra 
were compared to a detailed Monte Carlo employing estimated ionization and scintillation yields for nuclear recoils. 
The ability to discriminate between electronic and nuclear recoils
using the ratio of ionization to primary scintillation is demonstrated.  
These results encourage further investigation on the use of xenon in the gas phase as a detector medium in dark matter direct detection experiments.
\end{abstract}

\maketitle

\section{Introduction\label{s_int}}
Xenon has been the detection medium of choice in multiple experiments searching for rare physics events due to its favorable properties as a detection
medium \cite{Noblegasdetectors,Aprile_2010} including the availability of two channels of energy measurement, scintillation and ionization, that can be accessed 
simultaneously in a single detector.  In particular, recent experiments have employed liquid xenon in searching for interactions of WIMP
(Weakly Interacting Massive Particle) dark matter \cite{LUX_2014, XENON100_2012, ZEPLINIII_2012}, and neutrinoless double-beta ($0\nu\beta\beta$) decay 
\cite{EXOPRL_2012}.  Both of these processes have strong implications in fundamental physics.  WIMPs are strong candidates to be a possible constituent of 
cold dark matter (see for example \cite{Bertone_2005}), thought to make up the majority of matter in the universe.  The observation of $0\nu\beta\beta$ 
decay (see for example \cite{Cadenas_2012}) would establish the Majorana nature of the neutrino and provide information on the absolute value of the 
neutrino masses and the neutrino mass hierarchy.

WIMPs interact via the electroweak force, allowing them to scatter off nuclei, and so the signature of a WIMP in a pure xenon detector would be the recoil
of a xenon nucleus, in which the energetic nucleus excites and ionizes xenon atoms to produce primary scintillation photons and
electron-ion pairs.  Nuclear recoils have been observed and well-characterized in liquid xenon.  In particular, it is known that the scintillation
and ionization yields of nuclear recoils are lower, or quenched, relative to those of energetic electrons (electronic recoils) of the same kinetic energy.  
A model that predicts these yields based on the existing measurements has been constructed in \cite{NEST_2011}.  Liquid xenon has also clearly shown that 
the amount of quenching in both scintillation and ionization is not the same, enabling one to
discriminate between electronic recoils and the nuclear recoil signals of interest to dark matter 
detection by using the ratio of ionization to primary scintillation (see for example \cite{Aprile_2007, Sorensen_2009, Horn_2011}).

The use of xenon in the gas phase may provide several advantages that would imply greater sensitivity in searches for dark matter and $0\nu\beta\beta$
decay.  In particular, the gas phase offers improved energy resolution \cite{Bolotnikov} over the liquid phase, largely believed to be due to the
observed significant fluctuation in energy deposition between the ionization and scintillation channels \cite{Conti} in liquid Xe.  Though this can be 
corrected by combining both channels to recover some of the lost energy resolution in the liquid phase, as is done in \cite{EXOPRL_2012}, the inability 
to achieve light collection efficiencies beyond $\sim 20\%$ (the LUX experiment \cite{LUX_2014} achieved an average detection efficiency of 14\% for primary
scintillation) limits overall resolution via Poisson fluctuations in the detection of a relatively small 
signal.  Better energy resolution could lead to improved electron/nuclear recoil discrimination.  Under the right conditions and possibly 
with the addition of a molecular additive to the pure xenon gas, the amount of electron-ion recombination in nuclear recoil tracks may depend on the 
orientation of the drift electric field, thereby providing information about the direction of a WIMP scatter \cite{Nygren_2013, Herrera_TIPP2014}.

Here we report data on the ionization and scintillation of nuclear recoils in gaseous xenon.  Further details of this study can be found
in \cite{Renner_thesis} and \cite{NEXT_TAUP2013}.  In addition, scintillation and ionization of nuclear recoils were previously presented in \cite{White_PPC}.
The experiment was performed with a high pressure xenon gas time projection chamber (TPC) constructed as a prototype for the NEXT (Neutrino Experiment with 
a Xenon TPC) experiment, the NEXT 
prototype for research and development towards detection of neutrinoless Double Beta and Dark Matter (NEXT-DBDM).  NEXT intends
to search for $0\nu\beta\beta$ decay with an electroluminescent TPC containing 100 kg of enriched (91\% $^{136}$Xe) xenon.  Should potential advantages
be found in using gaseous xenon to search for dark matter, one could comtemplate a simultaneous $0\nu\beta\beta$ and dark matter search with
a ton-scale gaseous xenon detector.  A clear understanding of nuclear recoils in gaseous xenon is a critical first step in this direction.
\section{Experimental Setup and Calibration\label{s_exp}}
\subsection{Detector Hardware and Operation}
\begin{figure*}
\includegraphics[scale=0.5]{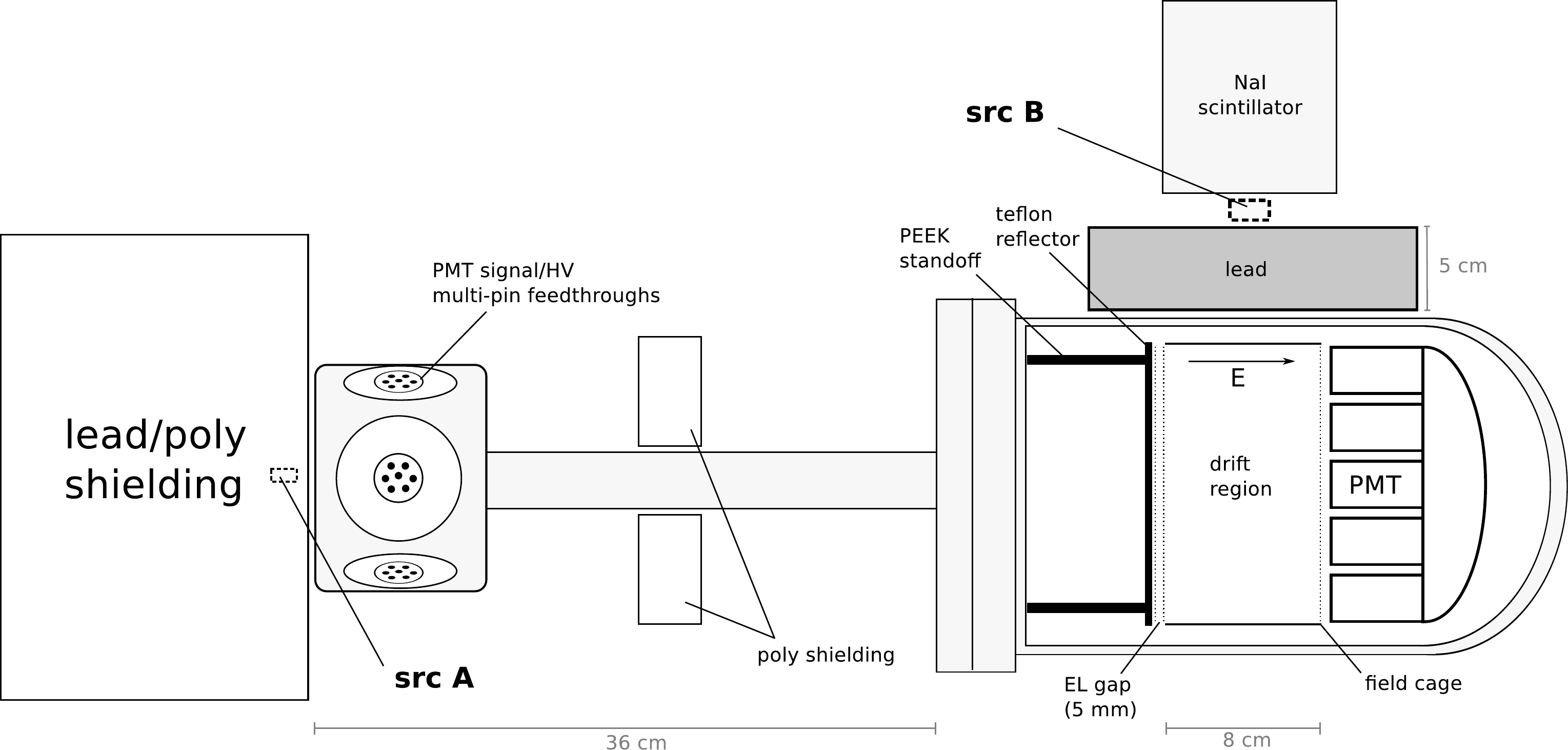}
\caption{\label{fig_exp_setup}Schematic of experimental configuration (not drawn to scale) for datasets used in this study.  When the source was at
position A, both the NaI scintillator and the source were enclosed in the lead/poly shielding.}
\end{figure*}

The NEXT-DBDM detector is described in detail in \cite{NEXT_2013}.  Here we summarize this description and describe the
modifications made for this study.  Figure \ref{fig_exp_setup} gives an overview of the experimental setup and source locations.

The main hardware of the TPC consists of a stainless steel cylindrical pressure vessel (20 cm diameter, 33.5 cm length) with one end closed in an 
ellipsoidal shape and the other sealed via a ConFlat flange to a stainless steel lid to which the internal components forming the TPC are attached.
The internal hardware consists of a hexagonal field cage separated into a drift (active) region of length 8 cm and an amplification region of length
5 mm by grids of wire mesh stretched tightly across metal frames.  The active region is enclosed by teflon panels with copper 
strips attached to their outer surfaces which are connected via resistors to grade the drift field.  The panels are supported by thin plastic frames, 
and the teflon surfaces facing the active region were coated with tetraphenyl 
butadiene (TPB) by dissolving the TPB in toluene 
and spraying it directly onto the surface using an airbrush.  An array of 19, 1-in.\ diameter Hamamatsu photomultiplier tubes (PMTs) arranged in a 
hexagonal pattern is located at the end opposite to the amplification region.  High voltages for the wire meshes are 
passed into the pressure vessel through the lid via commercial feedthroughs rated to 20 kV at 17 bar and connected via teflon-coated wire to the mesh frames.  
The lid is connected by a long tube to a stainless steel octagon with 8 ConFlat ports, several of which are occupied by multi-pin feedthroughs through
which PMT high voltages are input to the interior of the detector and through which the PMT signals are output.  An opening of diameter 1.7 cm extends 
through the center of the octagon and down the tube to a 2 mm source entrance window to the interior of the pressure vessel.  An external sodium iodide 
(NaI) scintillator coupled to a PMT was often used to detect gamma rays emitted in coincidence with the neutrons or gamma rays of interest.

The pressure vessel is connected to a gas system allowing for full system pump-down to pressures on the order of
$5\times 10^{-5}$ Torr.  The gas system also permitted reclamation/reintroduction of the xenon gas used in operation to/from a steel cylinder and constant 
recirculation of the xenon gas during operation through a heated zirconium-based getter, to remove impurities such as O$_{2}$, H$_{2}$O, and N$_{2}$.  
Typically after servicing the TPC, the entire system is pumped to $\sim 5\times 10^{-5}$ torr, flushed with Ar gas, pumped a second time, and filled with 
xenon gas.  The gas must be purified for several days before optimal electron lifetimes (of order 10s of milliseconds) are realized.

A typical event in the detector occurs when an energetic particle interacts in the xenon medium, producing primary scintillation in the form of 
$\lambda\sim 170$ nm photons, and ionization \cite{Noblegasdetectors}.  The primary scintillation is detected immediately by the PMTs and constitutes the signal denoted as S1.  The ionization
is drifted via an electric field to the amplification region, a narrow region of high field in which the electrons are accelerated to energies sufficient
to repeatedly excite but not ionize the xenon atoms in the medium.  Each excitation results in the emission of 
a xenon UV photon so that each individual electron traversing the high field region produces a number of photons $g$ equal to 
\cite{Nygren_2009, Monteiro_2007}

\begin{equation}
 g = 140(E/p - 0.83)p\Delta x,
\end{equation}

\noindent where $\Delta x$ is the thickness in cm of the region of high field $E$, given in kV/cm, and $p$ is the gas pressure in bar.  This process, called
electroluminescence, provides a means of amplification with relatively lower fluctuations than amplification based on electron avalanches
and results in a light signal denoted as S2 that is proportional to the number of ionization electrons produced in the event.

The TPB on the walls surrounding the active region shifts a significant fraction of the xenon scintillation light 
to the visible regime ($\lambda\sim 430$ nm) at which the quantum efficiency of the PMTs is higher.  Relative to results obtained before the TPB was
applied to the walls of the field cage, the TPB provided about a factor of 2 increase in 
light collection efficiency and was essential to observing the low S1 signals produced by nuclear recoils discussed in Sec.\ \ref{s_neu}.

All PMT signals are sampled continuously using a Struck SIS3302 digitizer and buffered in waveforms of 16384 samples that are stored in digitizer memory
when a trigger constructed from a network of NIM modules is activated.  The events are read out to a desktop computer in blocks of 512 and processed 
using an automated data management and analysis system based on ROOT \cite{ROOT} and FMWK \cite{FMWK}.  In the analysis, the signals from the 19 PMTs are 
baseline subtracted and summed, and peaks corresponding to PMT pulses are located and integrated to give a number of detected photons (see 
Fig.\ \ref{fig_neu_nwf_NaI} for an example waveform).  Each peak is considered as a candidate S1 or S2 pulse based on its width and, when the location of 
S1 in the waveform was fixed due to tagging with an external scintillator, its location.  Once a single S1 pulse and one or more S2 pulses have been 
identified, the drift time of the ionization produced in the event can be determined as the difference in time between the beginning of the S1 pulse and 
the centroid in time of the S2 pulse.  One or more S2 peaks are possible in a single event if the ionization track has multiple components
that arrive at the amplification region at distinct times.  This could occur, for example, if a xenon fluorescent x-ray were produced and traveled several cm before making
a distinct ionization track of its own.  Key quantities such as the integrated numbers of photons in the S1 and S2 signals, the drift time, and pulse
integration and timing information from the external NaI scintillator are all stored in a ROOT tree recorded on an event-by-event basis at the end of 
the analysis to facilitate access.

\subsection{Sources and Configuration}\label{ss_exp_srccfg}
The data discussed in this study were acquired using radiation emitted from four different radioactive sources - two producing only gamma radiation and two
radioisotope neutron sources producing energetic ($\sim$ 1-10 MeV) neutrons.  The neutron-producing $^{9}$Be$(\alpha,\mathrm{n})^{12}$C reaction that occurs
within the sources often leads to the emission of a high-energy gamma ray in coincidence with the emitted neutron.  This gamma ray has an energy of
either 4.4 MeV or 7.7 MeV.  To more efficiently isolate neutron-induced events in this study, the detection of a 4.4 MeV gamma ray using a NaI
scintillator was included as one of the trigger conditions when data was taken with a neutron source.  Further details on the neutron production mechanisms 
and spectra of the radioisotope neutron sources are discussed in detail in the Appendix.  

The configurations of the sources used in this study are itemized in the following subsections.  Of the 19 PMTs in the
energy plane, 18 were operational during the acquisition of all datasets, and all datasets were taken with a drift field of 370 V/cm at a gas pressure of 
approximately 14 bar.

\subsubsection{$^{241}$Am/Be Neutron Source}
An americium-beryllium neutron source containing a mass of $^{241}$Am with approximately 56 mCi of activity was positioned in source location A as shown in
Fig.\ \ref{fig_exp_setup}, just in front of the long tube connecting to the pressure vessel lid.  The source was surrounded by a layer of lead
about 2 in.\ thick and a layer of polyethylene also about 2 in.\ thick, such that it was only exposed in the direction along the tube.  Additional polyethylene
shielding was placed along the tube to collimate the incident neutron flux to the cross-sectional area of the tube.  The NaI scintillator
was placed within the shielding nearby the source to tag 4.4 MeV gamma rays emitted from it.  In this configuration, neutrons were emitted
far enough from the xenon volume that their interactions could be distinguished from those induced by gamma rays.  This was done by using the time-of-flight 
measured as the time difference in the arrival of the S1 produced in the TPC and the gamma ray tagged in the NaI scintillator.

\subsubsection{$^{238}$Pu/Be Neutron Source}
A plutonium-beryllium neutron source containing a mass of $^{238}$Pu with approximately 10 mCi of activity was positioned at the side of the pressure
vessel between the NaI scintillator and a lead brick at source location B of Fig.\ \ref{fig_exp_setup}.  The presence of the lead greatly reduced the
background due to gamma rays emitted as products of the neutron-generating $(\alpha,\mathrm{n})$ reaction.

\subsubsection{$^{22}$Na Gamma Source}
A sodium source containing a mass of $^{22}$Na with approximately 10 $\mu$Ci of activity was placed at the side of the pressure vessel, similar to
the configuration shown for source B in Fig.\ \ref{fig_exp_setup}, though no lead block was present, the source was positioned about 16 cm from the side 
of the pressure vessel, and the NaI scintillator was positioned several centimeters away from the source to avoid pileup.  Two collinear 511 keV gamma rays 
are emitted from the source.  One gamma ray was tagged with the NaI scintillator while the other was incident on the xenon volume.

\subsubsection{$^{137}$Cs Gamma Source}
A cesium source containing a mass of $^{137}$Cs with approximately 1 mCi of activity was contained in a lead enclosure to which a cylindrical lead
collimator with an opening of diameter 3.5 mm was fitted at one end.  The resulting collimated beam of 662 keV gamma rays was placed at source location A as
shown in Fig.\ \ref{fig_exp_setup} and pointed down the tube connecting the octagon and pressure vessel through the 2 mm thick source entrance window.

\subsection{Detector Calibration with a $^{137}$Cs Source}\label{ss_exp_erecoils}
Before discussing the characterization of nuclear recoils, we establish a gamma-based calibration of the detector using a $^{137}$Cs source 
(configuration 4 of Sec.\ \ref{ss_exp_srccfg}).  Gamma rays of energy 662 keV were directed axially through the center of the TPC lid, and the S1 and S2
signals were examined to determine key xenon properties such as the amount of energy required to produce a primary scintillation photon $W_{\mathrm{sc}}$, 
the relation used to correct for the z-dependence (lower detection probability of photons produced farther from the PMT array) of S1, and the light 
collection efficiency $\epsilon$ at the EL plane.  We first assume a value for the amount of energy required to produce an ionization electron 
$W_{\mathrm{i}} = 24.7 \pm 1.1$ eV (see Sec.\ 3.2 of \cite{Nygren_2009}).

\begin{figure}
\includegraphics[scale=0.48]{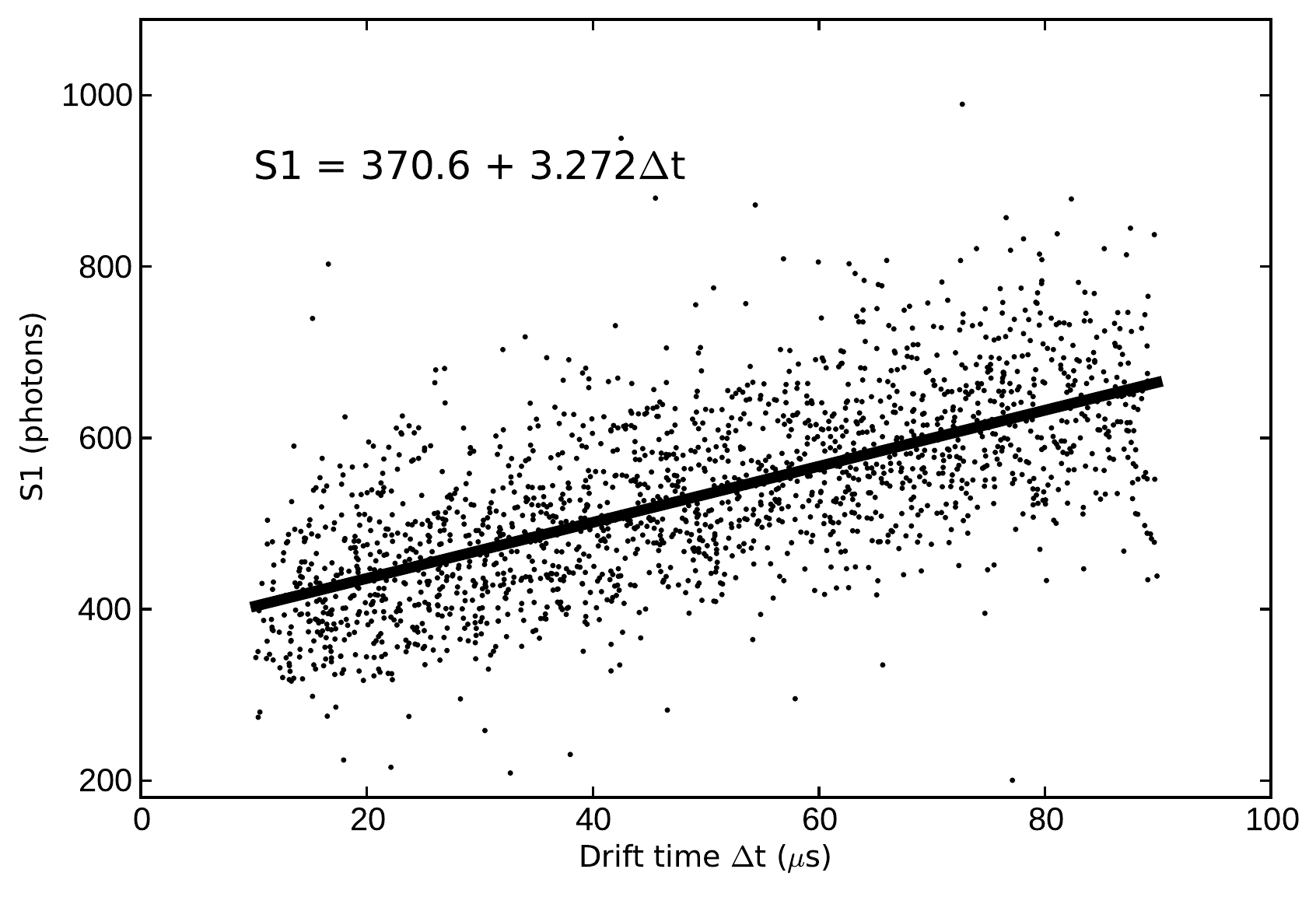}
\caption{\label{fig_exp_s1vstime}S1 vs.\,drift time (effectively the z-coordinate) for events determined to fall in the 662 keV peak in a $^{137}$Cs 
calibration run.  The linear fit was used to determine a correction factor to compute the S1 value for each event as if it occurred at the EL plane (z = 0).}
\end{figure}

Events corresponding to full-energy depositions of 662 keV gamma rays were isolated as a peak in the S2 distribution produced by the source.  A central
fiducial cut was made according to the weighted average (x, y) location of the event determined using the distribution of electroluminescent light produced 
on the PMT plane, and the events were corrected for electron attachment by multiplying by a z-dependent exponential factor corresponding to an electron 
lifetime of $\tau_{e} \approx 8.3$ ms.  The $^{137}$Cs S2 photopeak was found to lie at $S_{2} = 561617 \pm 92$ photons.  Operating at EL gain 
$g = 734 \pm 89$ photons/e$^{-}$, we calculate for the light collection efficiency at the EL plane,

\begin{equation}\label{eqn_exp_eps}
 \epsilon = \frac{S_{2}\cdot W_{\mathrm{i}}}{g\cdot E_{\gamma}} = 0.0285 \pm 0.0037,
\end{equation}

\noindent where $E_{\gamma} = 661657 \pm 3$ eV \cite{LBLTOI} is the energy of the gamma ray.  The geometrical z-dependence of S1 detection efficiency 
was determined by plotting the integrated S1 values of the events in the $^{137}$Cs S2 photopeak vs.\,event drift time 
(see Fig.\ \ref{fig_exp_s1vstime}).  A linear dependence is observed,

\begin{equation}\label{eqn_exp_S1fit}
 S_{1} = S_{1,0} + k\Delta t,
\end{equation}

\noindent where $\Delta t$ is the drift time and $S_{1,0}$ is a constant corresponding to the number of S1 photons detected at the EL plane for
the 662 keV energy deposition.  It was determined that $S_{1,0} = 370.6 \pm 4.0$ 
photons and $k = 3.272 \pm 0.076 \,\, \mathrm{photons/\mu s}$, so that the S1 signal detected for an event with drift time $\Delta t$, denoted
as $S_{1}(\Delta t)$, could be corrected for its z-dependence as

\begin{equation}\label{eqn_exp_S1corr}
 S_{1}' \equiv S_{1,0}\cdot\frac{S_{1}(\Delta t)}{S_{1,0} + k\Delta t} = \frac{S_{1}(\Delta t)}{1+(k/S_{1,0})\Delta t}.
\end{equation}

Knowing how to correct the S1 signal to its value at the EL plane, and knowing the light collection efficiency at the EL plane, we can compute the energy
required to produce an S1 photon $W_{\mathrm{sc}} = (E_{\gamma}/S_{1})\epsilon'$, where $\epsilon' = \epsilon/\Omega$ and $\Omega$ is a factor accounting for
the optical effect of the EL wire mesh grids.  This correction is necessary because $\epsilon$ is calculated using S2 light produced in the EL gap
between two grids while the S1 light is produced in the active region.  The factor $\Omega$ was determined by a Monte Carlo simulation
in which photons were generated within the EL gap and just in front of the EL gap, and the resulting efficiencies in the two cases were 
compared.  The grids in the simulation were given the nominal transparency of the physical mesh grids, equal to 88\% at 0\textdegree\,incidence angle, and
from the Monte Carlo, the relative light collection efficiency was determined to be described by $\Omega = 0.83\pm0.08$, assuming 10\% errors.  Using 
Eq.\ \ref{eqn_exp_eps}, we find

\begin{equation}\label{eqn_exp_Wsc}
 W_{\mathrm{sc}} = \frac{S_{2}\cdot W_{\mathrm{i}}}{S_{1}\cdot g\cdot \Omega} = 61.4\pm 18.0 \,\, \mathrm{eV},
\end{equation}

\noindent with an applied drift field of 370 V/cm, and note we have assumed an additional systematic error of 15 eV obtained from Monte Carlo
studies (see Sec.\ \ref{ss_neu_yields}).  Because the Monte Carlo study attempted to construct a consistent picture from electronic recoil yields, EL gain,
and optical/geometrical effects in the TPC, this additional error could originate from discrepancies in any of these areas.  The obtained 
$W_{\mathrm{sc}}$ is lower than those obtained in other references including $W_{\mathrm{sc}} = 76 \pm 12$ eV \cite{Parsons_1990}
and $W_{\mathrm{sc}} = 111 \pm 16$ eV \cite{Carmo_2008}, though these experiments were carried out under significantly different operating conditions
and in one case \cite{Parsons_1990} in a mixture of 90\% Xe / 10\% He gas.

We also note that both S1 and S2 pulses appeared to possess long tails, a property which was not present during previous operation before TPB 
was placed on the walls of the field cage.  TPB itself is not expected to produce such an effect, and 
the present effect may be due to an agent present in the toluene-based solution employed in the coating process.  While the S2 light is produced over a 
timespan of several microseconds to tens of microseconds, the S1 light arrives over a short timescale, allowing for the characterization of the tail 
accompanying a single, fast pulse of light.  The S1 pulses were found to be well described by a two component exponential, one with a short decay time 
constant of $\tau_{s} \approx 100$ ns, and one with a longer decay constant of $\tau_{l} \approx 1.4 \,\mu$s.
\section{Nuclear Recoils in High Pressure Xenon Gas\label{s_neu}}
\subsection{Spectrum of Emitted Neutrons}\label{ss_nrspec}

\begin{figure}
\includegraphics[scale=0.48]{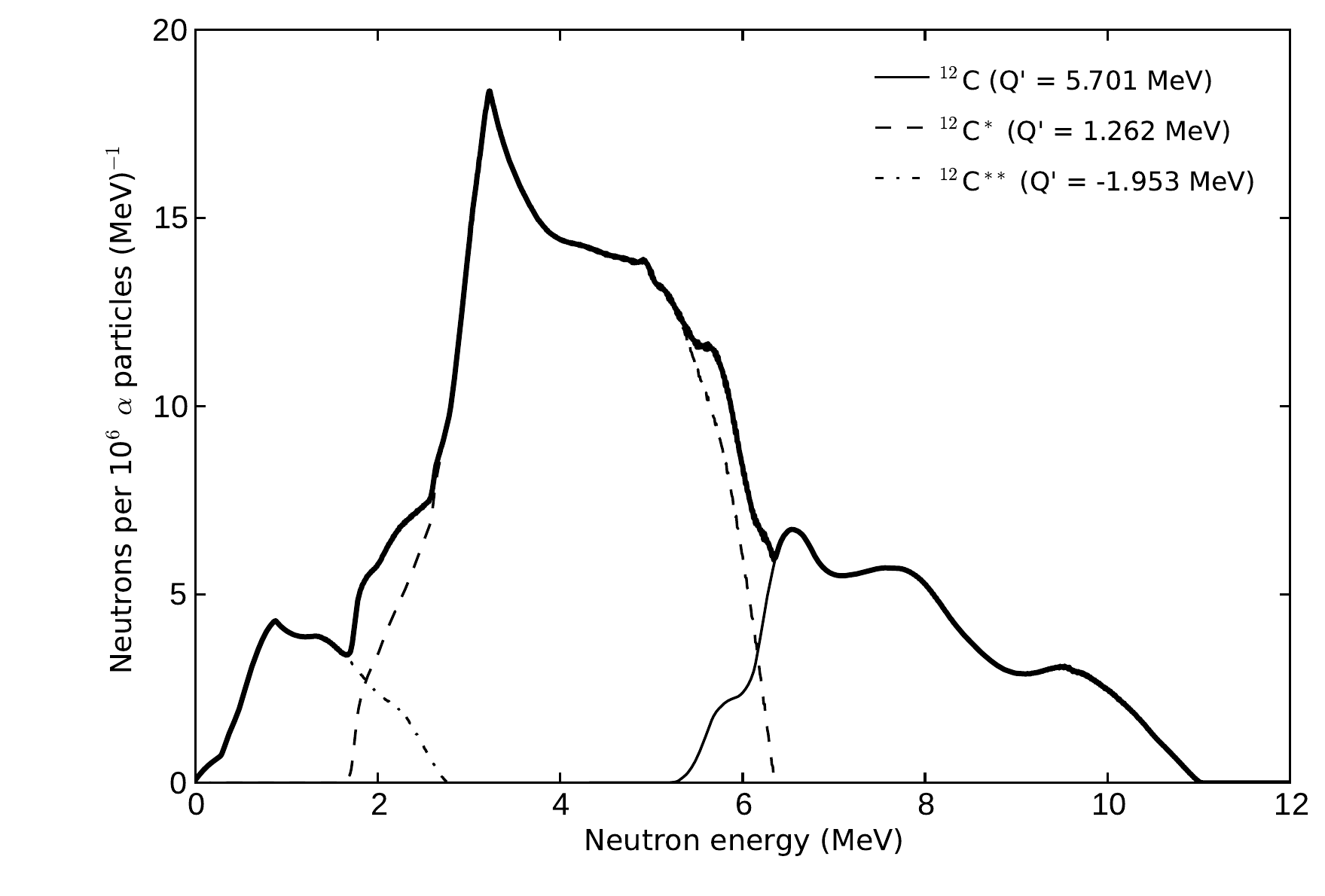}
\caption{\label{fig_neu_alnspectrum}Calculated neutron emission spectrum for a $^{241}$Am/Be neutron source considering only the neutrons
produced in the $^{9}$Be$(\alpha,\mathrm{n})^{12}$C reaction.  The spectrum is divided into three components for which the resulting carbon nucleus is
found in the ground (solid line), first excited (dashed line), or second excited (dot-dashed line) state.  The bold solid line shows the sum of all three
components.  The neutrons of greatest interest to this study are those for which the carbon nucleus is left in the first excited state and decays emitting
a 4.4 MeV gamma ray.}
\end{figure}

In this section, we present data taken with the radioisotope neutron sources, and we first show the expected spectrum of emitted 
neutrons from such sources.  The neutron spectrum from a $^{241}$Am/Be source is shown in Fig.\ \ref{fig_neu_alnspectrum} and was calculated assuming the 
source consisted of a volume filled with beryllium throughout which the isotope $^{241}$Am was uniformly distributed.  Further details and a description of 
the relevant calculations can be found in the Appendix.

\subsection{Analysis of Experimental Data}\label{ss_neu_analysis}
The neutrons emitted from the radioisotope sources used may be accompanied by gamma rays, and often the neutrons themselves scatter inelastically off
nuclei in the environment (in the xenon or surrounding detector hardware), resulting in the emission of various gamma rays in the de-excitation of the nuclei.
Therefore, careful analysis is necessary to isolate the nuclear recoil events, and to do so, we make a sequence of analysis cuts on the data to 
remove gamma-induced events and finally identify a band of events at low values of S1 and S2 that consists primarily of nuclear recoils.

The data presented here, consisting of 3682304 events in total, was 
acquired over about 577 hours, during which time the acquisition system was actively recording events at a rate of about 3 Hz for about 346 hours.  The
rate of acquisition was stable throughout the majority of the run, and 6 blocks of 512 events were removed due to anomalous rates.  Most 
of the downtime was due to the readout and storage of events from the digitizer.  Events occurring within 5 minutes of occasional high voltage breakdown 
within the detector were discarded.  This eliminated about 3.8\% of the events, thus slightly lowering the effective live-time.

The analysis cuts were performed in a sequence consisting of: tagging and single-pulse event identification, time-of-flight, diffusion, and radial
cuts.

\begin{figure}
\includegraphics[scale=0.48]{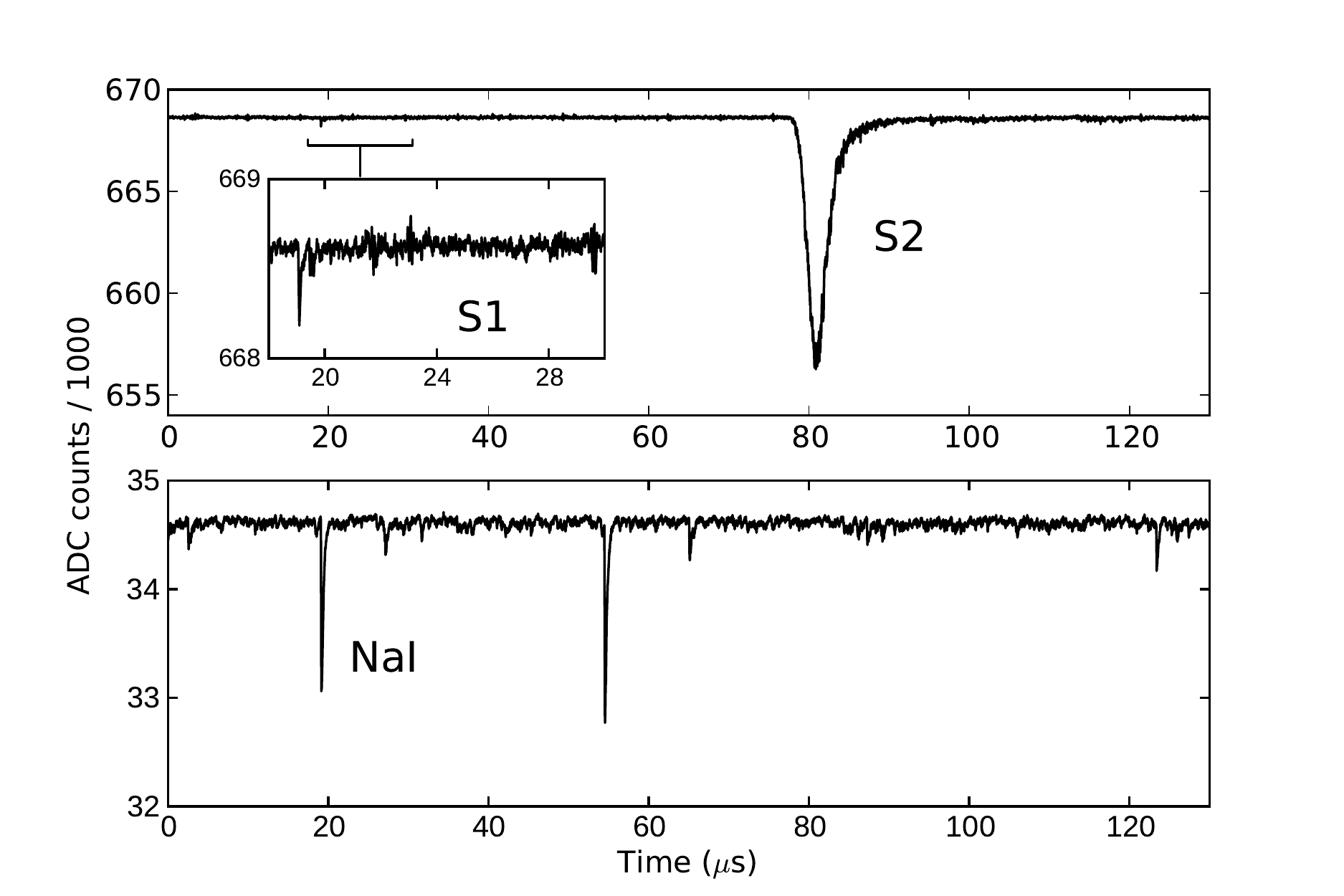}
\caption{\label{fig_neu_nwf_NaI}A typical candidate neutron event.  Note the S1 in coincidence with a pulse in the PMT coupled to the NaI scintillator,
followed by the single, Gaussian-like S2 pulse indicative of a pointlike energy distribution in the TPC.  The slight bulge in the right tail of the S2 pulse
is due to extended light emission from the presence of the TPB coating on the teflon walls of the field cage.}
\end{figure}

A typical tagged neutron candidate event is shown in Fig.\ \ref{fig_neu_nwf_NaI}.  The S1 signal and NaI pulses were fit to a function of the form

\begin{equation}\label{eqn_neu_S1fit}
 f(x) = p_{0} + \frac{p_{1}e^{-(x-p_{2})/p_{4}}}{1 + e^{-(x-p_{2})/p_{3}}},
\end{equation}

\noindent where $x$ is the sample value and the $p_{i}$ are fit parameters.  The value of $p_{2}$ for each type of pulse was taken to be its initial 
arrival time.  An event was required to contain at least one pulse in the NaI scintillator and an S1 pulse, both with arrival times in a sample region near 
the expected S1 arrival time defined by the trigger.  The integrated area of an NaI pulse was required to lie within a broad region chosen to correspond 
to the spectrum of a 4.4 MeV
gamma deposition.  If multiple NaI pulses met the required criteria on arrival time and charge, the NaI pulse closest in start time to that of the S1 signal 
was selected.  The event was also required to contain a single S2 pulse, that is, one integrated pulse as defined by the pulse-finding and integration
algorithm with charge $q_{2}$ nearly equal to the total integrated S2 charge $Q_{2}$, more specifically $0.95q_{2} < Q_{2} < 1.05q_{2}$.  An event for which
all of these conditions are met is considered to have passed the tagging and single-pulse identification cuts.

\begin{figure}
\includegraphics[scale=0.48]{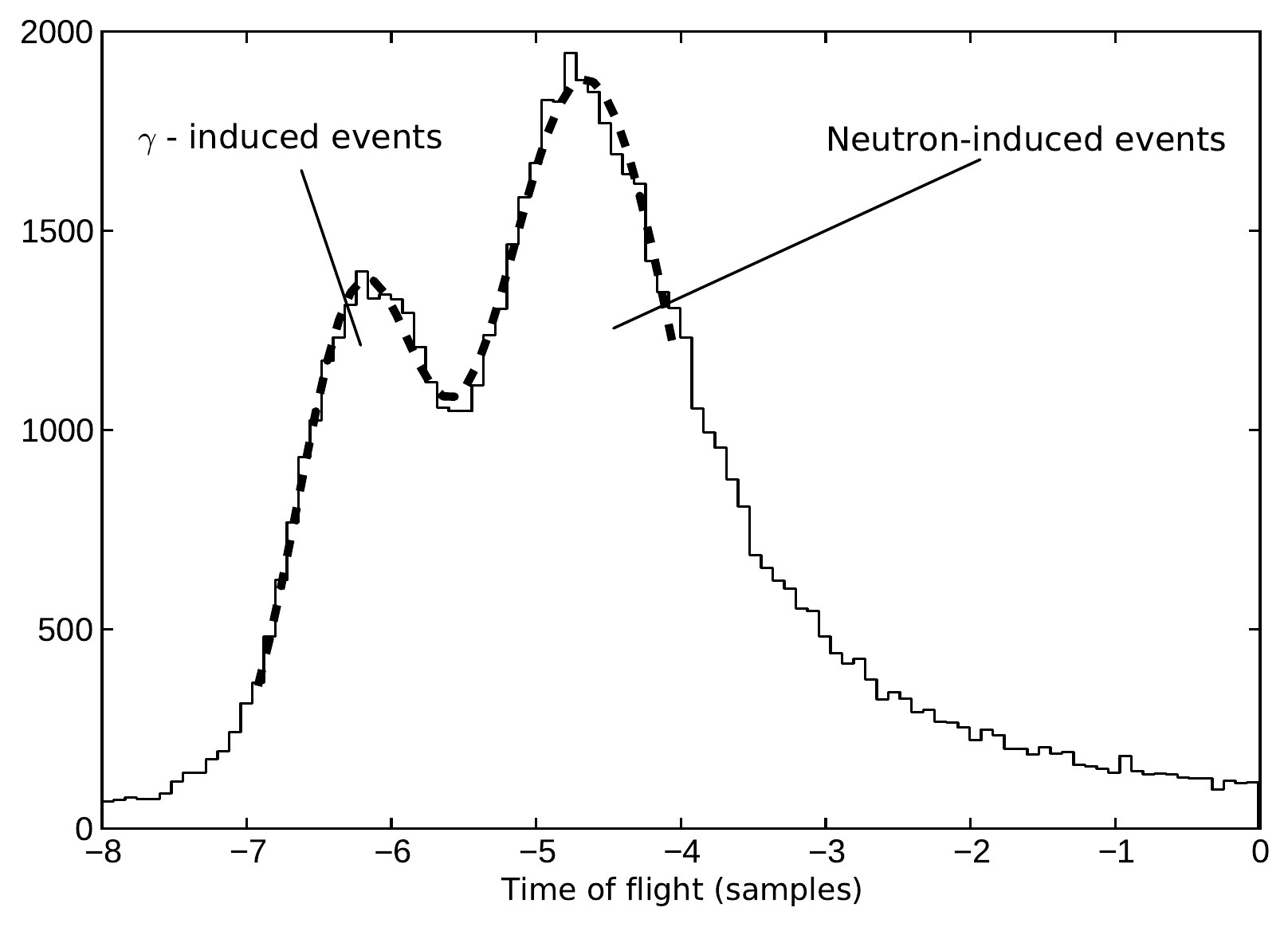}
\caption{\label{fig_neu_tof_bands}Time-of-flight distribution (1 sample = 10 ns) for properly tagged events with S2 $>$ 175000 photons and 30 photons $<$ S1 $<$ 500 photons 
(corrected for z-dependence).  The left peak in the distribution is due to gammas originating from the source, and the right peak is due to events induced by neutrons from
the source.  The fit to a sum of two Gaussians is shown for which $\mu_{1} = -6.24$, $\sigma_{1} = 0.429$ (left) and $\mu_{2} = -4.66$, $\sigma_{2} = 0.647$ 
(right).}
\end{figure}

The time-of-flight $\tau$ for a given event is determined for properly tagged events as the difference between the arrival times of the S1 and NaI pulses and is
given by

\begin{equation}\label{eqn_neu_tof}
 \tau = \biggl[\frac{1+E_{0}/K}{\sqrt{1+2E_{0}/K}}\biggr](d/c) + \Delta t_{\mathrm{off}}
\end{equation}

\noindent where $E_{0}$ and $K$ are the rest and kinetic energies of the particle, $d \approx 50$ cm is the distance traveled by the particle from the source to the point 
of detection, $c$ is the speed of light, and $\Delta t_{\mathrm{off}}$ is an offset time due to the trigger and electronics.  The tagging procedure serves to 
eliminate a significant amount of gamma background, and a clear peak in time-of-flight due to events induced by gammas from the source is not evident
in the overall time-of-flight distribution.  However, by examining only higher energy depositions consisting of a relatively higher fraction of gamma-induced
events, two peaks are apparent in the time-of-flight distribution.
Figure \ref{fig_neu_tof_bands} shows the relevant regime of the time-of-flight distribution for high-S2 events, in which the leftmost peak corresponds to events produced by gammas
emitted by the source.  The fit shown to a sum of two Gaussians gives for the component corresponding to the left peak, a mean of 
$\mu_{\gamma} \equiv \mu_{1} = -6.24$ and sigma $\sigma_{\gamma} \equiv \sigma_{1} = 0.429$.  We can then solve using Eq.\ \ref{eqn_neu_tof} with $E_{0} = 0$, $c \approx 30$ cm/ns and $\tau = \mu_{\gamma}$ for 
$\Delta t_{\mathrm{off}} = \mu_{\gamma} - d/c \approx -64$ ns.  Since for an active region of length 8 cm all gammas emitted from the source should arrive 
within a time interval of 8 cm / c = 0.27 ns, which is significantly less than $\sigma_{\gamma} = 4.29$ ns, we can use $\sigma_{\gamma}$ as a measure 
of the time-of-flight resolution, noting that for nuclear recoil 
events we expect the resolution to be poorer due to their lower S1 signals.  These S1 signals may consist of only several photons arriving within a time 
interval relatively large compared to the duration of a single photoelectron signal in a PMT, thereby producing inconsistencies in the fit to 
Eq.\ \ref{eqn_neu_S1fit}.

\begin{figure}
\includegraphics[scale=0.48]{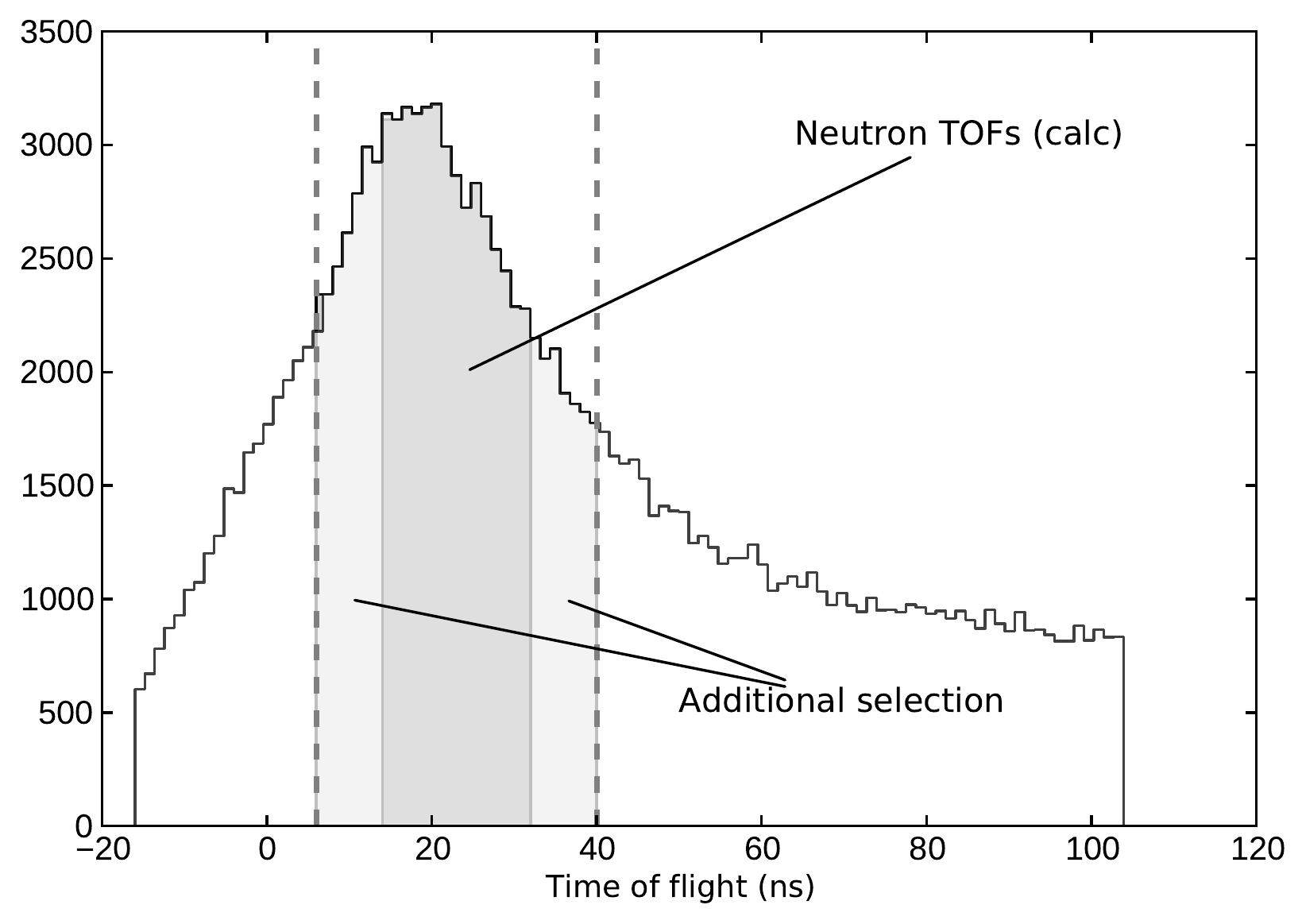}
\caption{\label{fig_neu_tof}Time-of-flight distribution for properly tagged events with S2 $<$ 40000 photons and S1 $<$ 50 photons 
(corrected for z-dependence).  The shaded region shows the time-of-flight cuts applied to select 6 ns $< \tau < $ 40 ns.}
\end{figure}

Figure \ref{fig_neu_tof} shows the time-of-flight distribution for low-energy S2 events.  Here the time-of-flight has been calculated in nanoseconds and 
the offset $\Delta t_{\mathrm{off}}$ removed.  Cuts were made according to the maximum and minimum times-of-flight generated in a Monte Carlo using
the calculated neutron source spectrum described in Sec.\ \ref{ss_nrspec}, allowing for an additional 8 ns (approximately $2\sigma_{\gamma}$) on both ends of 
the cut range.  Any event with time-of-flight within the selected region shown in Fig.\ \ref{fig_neu_tof} was considered to have passed the time-of-flight
cuts.

\begin{figure}
\includegraphics[scale=0.48]{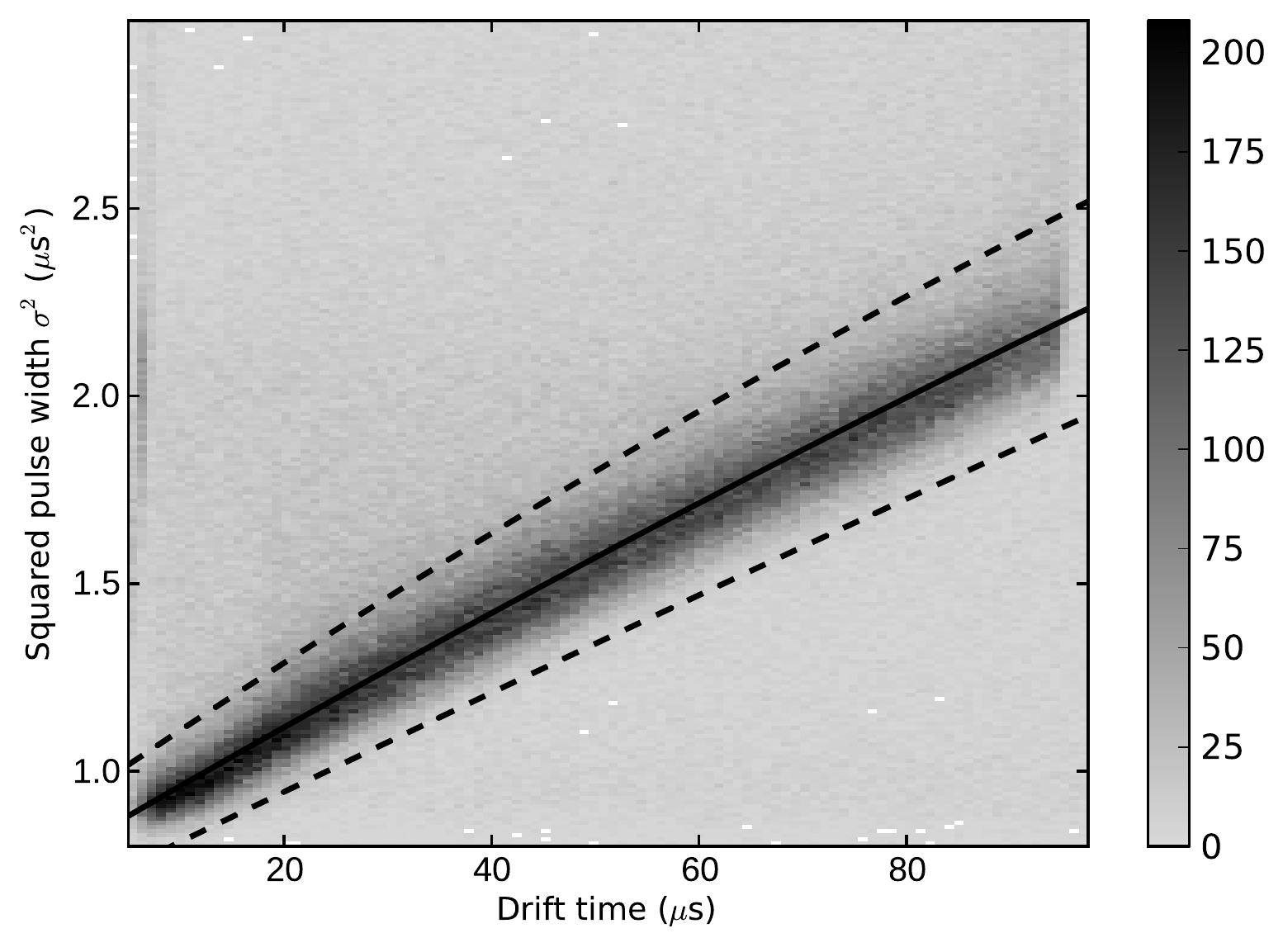}
\caption{\label{fig_neu_diff}Pulse width $\sigma^{2}$ from a Gaussian fit to the S2 pulse in each event plotted against drift time for events near and
within the diffusion band.  The centroid of the band is marked with a solid line, and dashed lines define the region of $\pm 2$ standard deviations
from the centroid.  The equation of the centroid fit shown is $\sigma^{2} = 0.801 + 0.0161\bar{t} - 0.0000143\bar{t}^{2}$.}
\end{figure}

Electron diffusion provides an additional means of confirming the validity of an event with a single S2 pulse.  If the selected S1 truly corresponds to
the primary scintillation produced by the ionization collected as S2, the width of the S2 pulse
will increase with the drift time of the ionization approximately as \cite{NEXT_alphas}

\begin{equation}\label{eqn_neu_sigmasq}
 \sigma^{2} = \sigma_{0}^{2} + 10D_{L}^{2}\bar{t}/v_{d},
\end{equation}

\noindent where $\sigma_{0}^{2}$ is a constant determined by the drift time of the electron across the electroluminescent (EL) gap, $D_{L}$ is the 
longitudinal diffusion constant in 
mm/$\sqrt{\mathrm{cm}}$, $v_{d}$ is the electron drift velocity in mm/$\mu$s, and $\bar{t}$ is the drift time in $\mu$s.  Note that this relation is 
an approximation as it assumes a Gaussian S2 pulse, which may not be strictly the case for events originating near the EL gap.  Events for which S1 was
properly selected will fall in a band described approximately by Eq.\ \ref{eqn_neu_sigmasq}, shown in Fig.\ \ref{fig_neu_diff}, in which the S2
pulse width was determined by a Gaussian fit.  The centroid and $\pm 2\sigma$ lines of the band were determined using a procedure similar to the one applied
in \cite{Dahl}.  Using a maximum drift time of about 95 $\mu$s and a drift length of 8 cm, we have $v_{d} = 0.84$ mm/$\mu$s, and using the constant and linear
terms of the centroid fit line, we find $\sigma_{0} = 0.89$ $\mu$s and $D_{L} = 0.37$ mm/$\sqrt{\mathrm{cm}}$.  All events within the $\pm 2\sigma$ lines 
are considered to have passed the diffusion cuts.

\begin{figure}
\includegraphics[scale=0.48]{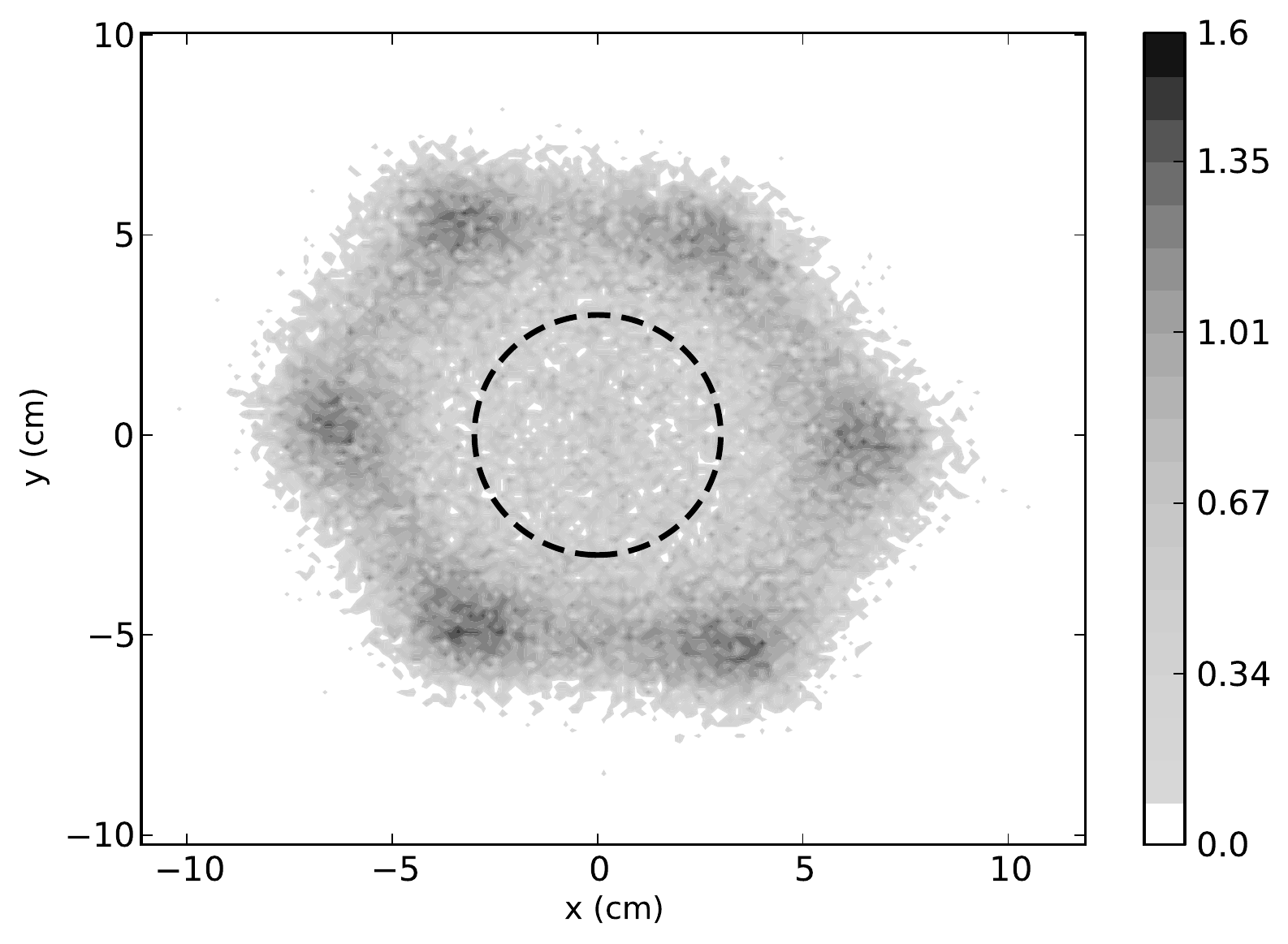}
\caption{\label{fig_neu_radcut}The distribution of average reconstructed (x,y) locations for events with S2 $<$ 40000 photons.  The shading is done on
a $\log_{10}$ scale.  The dashed circle defines a fiducial cut of $r < 3$ cm.}
\end{figure}

Using the pattern of light cast upon the PMT plane during S2 production, an average (x,y) location for each event can be constructed.  Though this method
does not provide the precision of a finer-granularity tracking plane, it allows for some fiducialization and, therefore, elimination of events that originate
from the walls of the TPC.  The average x and y coordinates are calculated by weighting the position of each PMT by the amount of S2 signal observed by
that PMT.  The resulting pattern is shown in Fig.\ \ref{fig_neu_radcut} and has been scaled and shifted so that it is centered upon (0,0) and its dimensions match those of
the physical dimensions of the PMT plane.  This required a shift of all reconstructed points by $(0.21,-0.20)$, scaling in x by a factor of 8.71, and scaling
in y by a factor of 12.6.  The scaling and shift procedure was necessary due to the uniformity of the light pattern cast on the PMTs and uncertainties in 
the individual PMT single-photon responses.  A fiducial cut of $r < 3$ cm is superimposed, and events lying inside the selected region pass the cut.  The 
tight fiducial cut is made here to show more clearly where the nuclear recoils lie in S1-S2 space, but this cut is varied in the forthcoming
discussions to obtain increased statistics at the expense of more background events.

\begin{figure}
\includegraphics[scale=0.45]{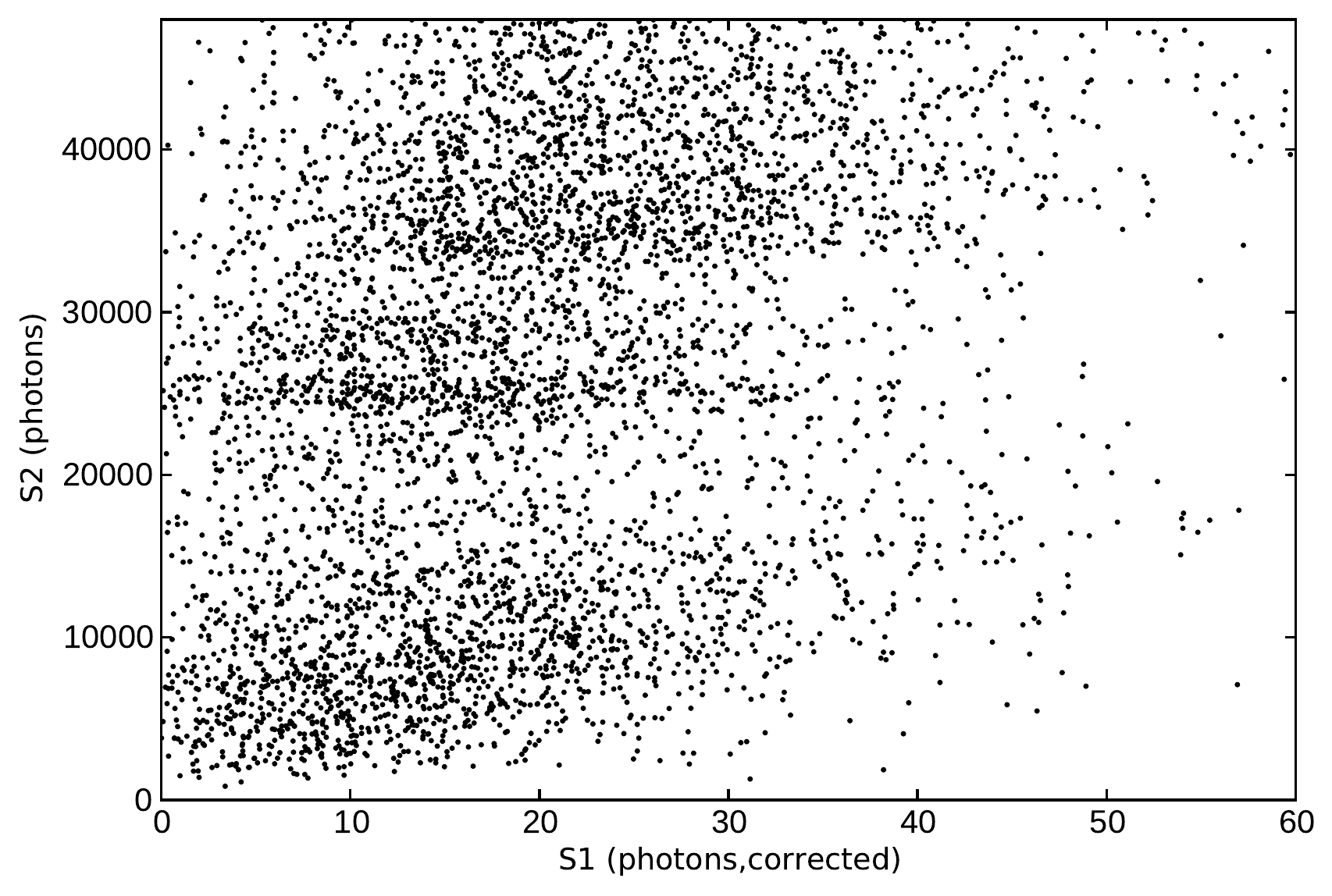}
\includegraphics[scale=0.48]{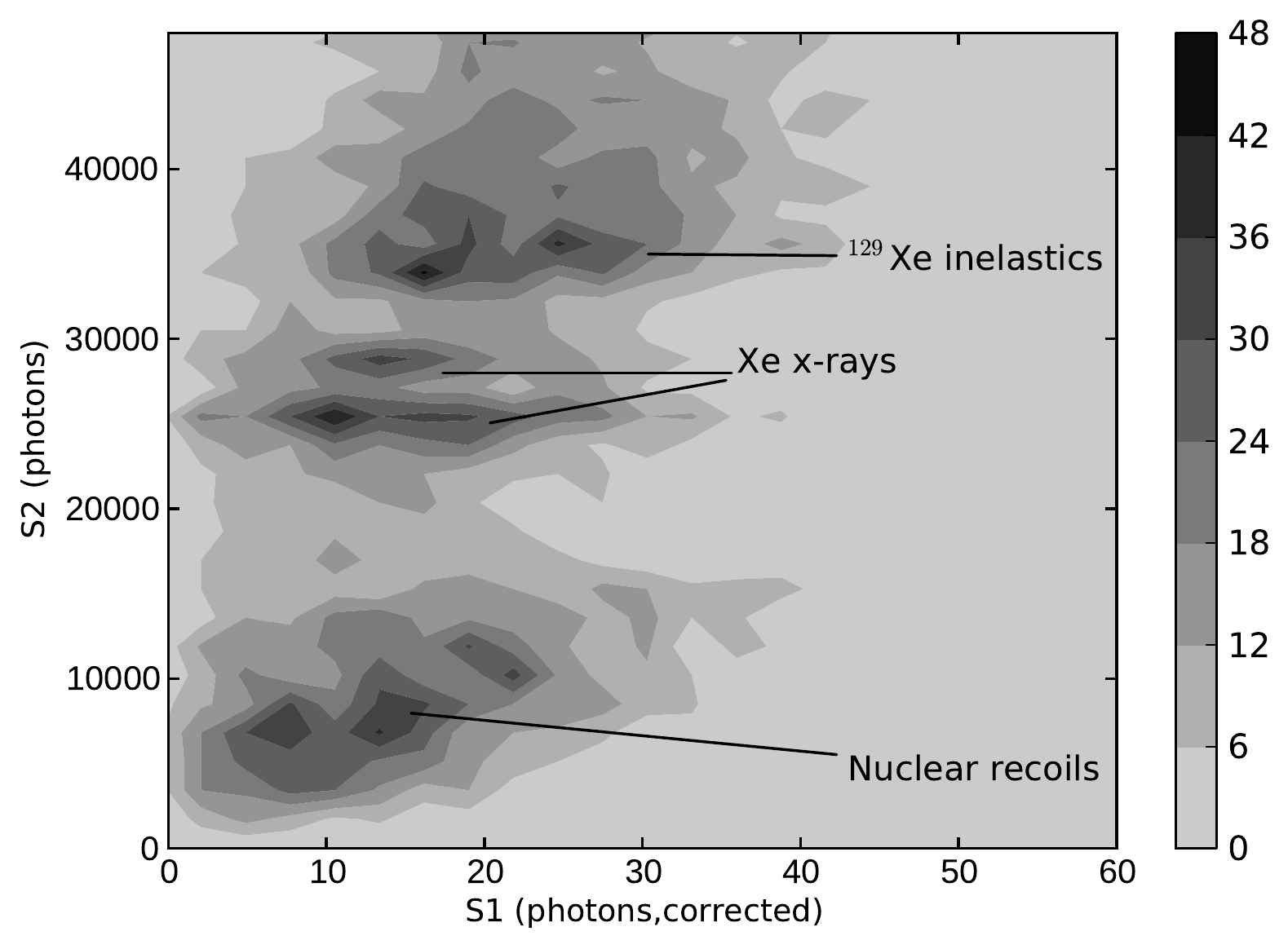}
\caption{\label{fig_neu_s1s2band}S1 (corrected for z-dependence) vs. S2 signals for events passing tagging, single-pulse, time-of-flight, diffusion, and 
radial cuts, shown in scatter (above) and contour (below) formats.  Events produced by neutron inelastic scattering on $^{129}$Xe (approx. 40 keV), by xenon 
fluorescent x-rays (approx. 29 keV and 34 keV), and by nuclear recoils lie in distinct bands on the plot.}
\end{figure}

From the known characteristics of nuclear recoil signals in liquid xenon, one suspects a class of events with low S1 and S2
and a different S2/S1 slope than the electronic recoil events.  Figure \ref{fig_neu_s1s2band} shows the relevant region of (S1, S2) space, including
the low-energy nuclear recoil band, for all tagged events passing the single-pulse, time-of-flight, diffusion, and radial cuts ($r < 3$ cm) described above.

\subsection{Electronic and Nuclear Recoil Discrimination}\label{ss_neu_eNR}
\begin{figure*}[!htb]
\noindent\includegraphics[scale=0.48]{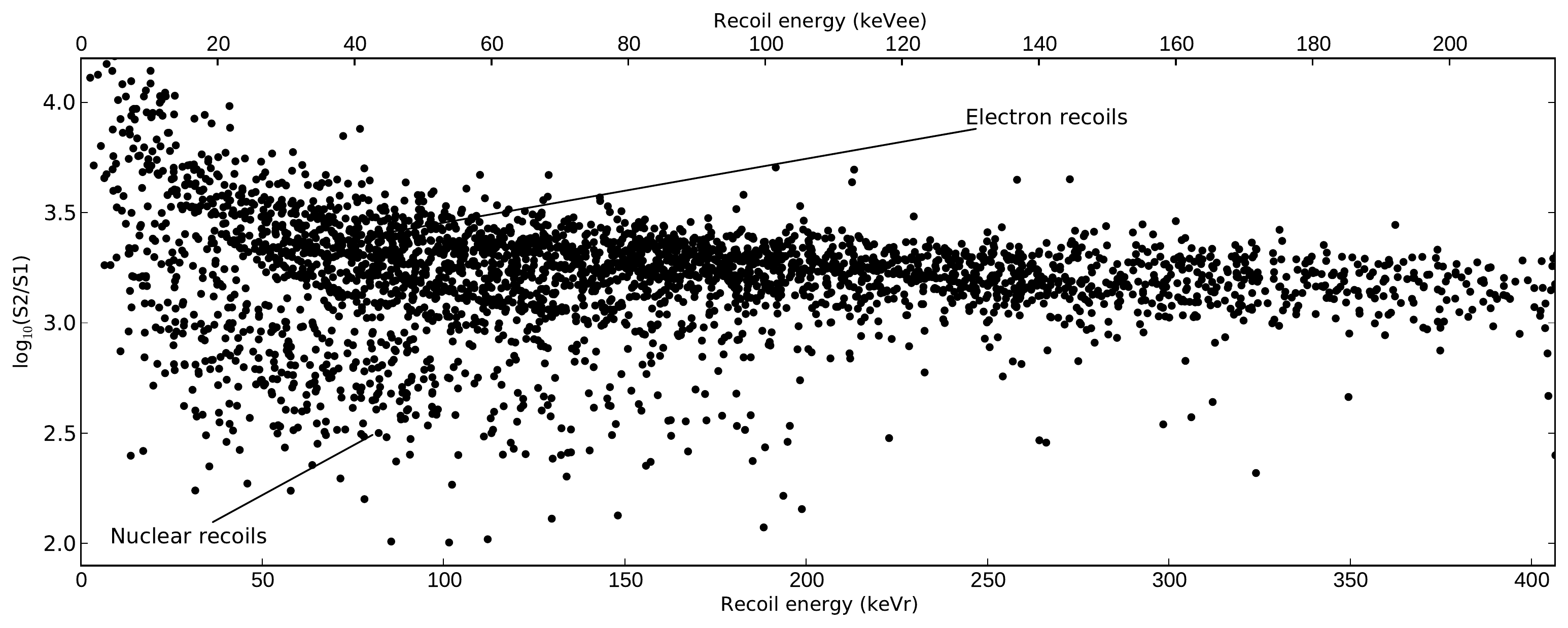}
\includegraphics[scale=0.48]{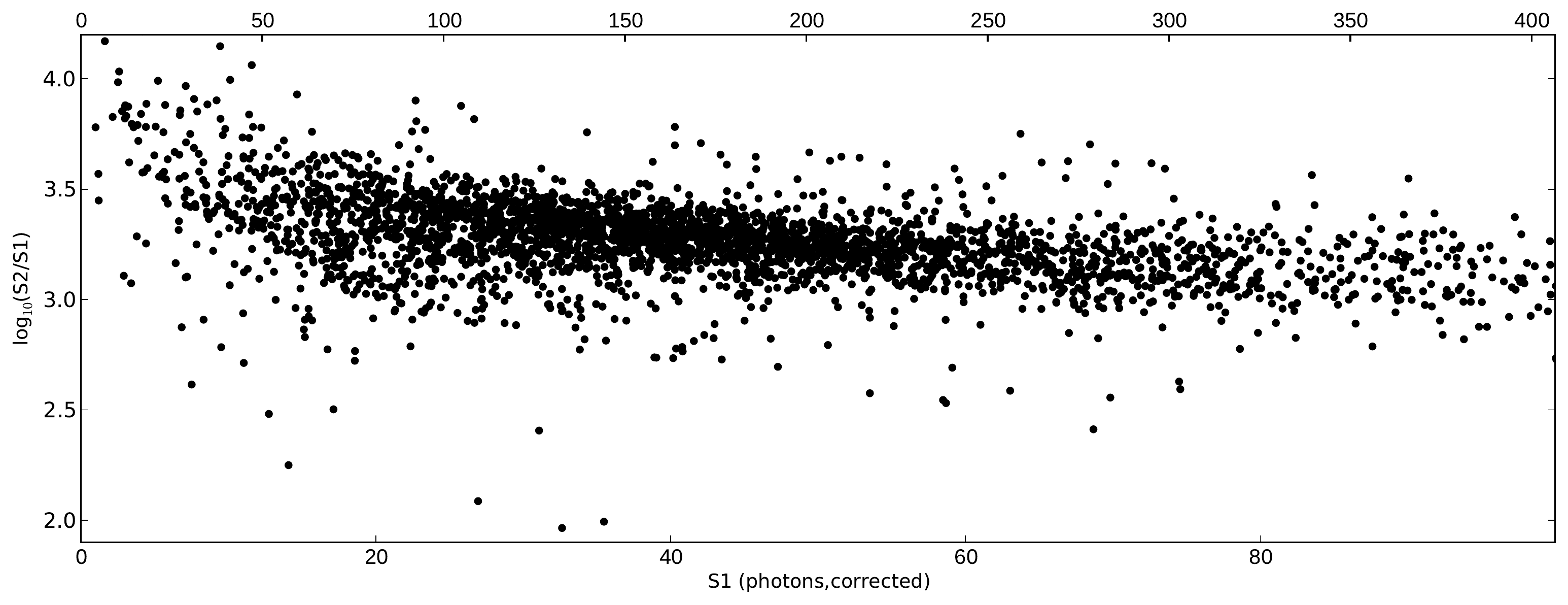}
\caption{\label{fig_neu_logs2s1c}The logarithm of the ratio of S2/S1 signals plotted (with equivalent scales) against S1 corrected for z-dependence for 
data taken with the $^{241}$Am/Be source (above) and a $^{22}$Na source (below).  The S1 values are calibrated to their corresponding electron recoil
energies (keV) and nuclear recoil energies (keVr) using the W-values $W_{\mathrm{i}}$ and $W_{\mathrm{sc}}$, EL gain $g$, and light collection
efficiency $\epsilon$ from section \ref{ss_exp_erecoils} and the estimated nuclear recoil quenching factors $\alpha_{\mathrm{S1}}$ and $\alpha_{\mathrm{S2}}$
from section \ref{ss_neu_yields}.}
\end{figure*}

For the purpose of dark matter searches, one is especially interested in the ability to discriminate between nuclear recoils (the potential signals) and
electronic recoils (background).  In liquid xenon, this can be done by examining the ratio of the S2/S1 signals produced in an event, as this ratio is
significantly lower on average for nuclear recoils.  Because the light collection efficiency in the experiments producing the present data is low 
(of order 3-5\%) and, therefore, the observed nuclear recoil events occupy a region of low S1 and S2 in which large fluctuations exist and the detection 
efficiency $\ne 1$, these data are not ideal for characterizing the full potential of the discrimination power of xenon gas.  However, they can be used to 
demonstrate that S2/S1 discrimination is possible and has potential in the gas phase.

Figure \ref{fig_neu_logs2s1c} shows the logarithm of the ratio S2/S1 plotted against S1 for data taken with the $^{241}$Am/Be neutron source and with
a $^{22}$Na source (configuration 3 of Sec.\ \ref{ss_exp_srccfg}).  The events shown passed the single-pulse, time-of-flight, 
diffusion, and fiducial cuts discussed in Sec.\ \ref{ss_neu_analysis}.  The number of events considered from the $^{241}$Am/Be dataset was reduced so that
the two datasets could be compared with similar statistics.
S1-S2 selection cuts were not applied so that events from both electronic and nuclear recoils could be shown.  The nuclear recoil events form a band 
clearly distinguishable from the electronic recoil events, though some background events lie in the nuclear recoil band in the $^{22}$Na data.  
This is due to several reasons, one being that at low S1, the small number of photons detected are subject to more significant Poisson fluctuations,
resulting in poorer resolution and greater likelihood of yielding an abnormally large S1 for a given S2.  Furthermore, it is possible to produce S1 in a 
region of the TPC from which S2 cannot be collected, for example in the small
gap between the PMT plane and the wire mesh that defines the beginning of the drift region.  Often a low-energy event
is accompanied by additional gamma rays that escape the active region, in which case an event with a single-pulse S2 could consist of additional gamma energy 
deposited but only seen in S1, yielding a range of possible S1 signals for a given S2.

\subsection{Monte Carlo and Estimated Ionization and Scintillation Yields}\label{ss_neu_yields}

The present results do not include information on the absolute energy of each nuclear recoil on an event-by-event basis, and so the ionization and
scintillation yields for nuclear recoils can only be determined in principle by using the measured recoil spectrum as a whole and comparing it to
expectations based on calculation and Monte Carlo simulation.  For example, the peaks at larger angles in the neutron elastic scattering cross section 
(see Appendix) should lead to the presence of an identifiable feature in the recoil energy spectrum near 80 keVr.
The presence of this feature is not statistically significant enough in the present data to make a strong definite claim, though its possible
presence is investigated in \cite{Renner_thesis} along with fits of experimental data to Monte Carlo spectra, and the results are
used to obtain estimated nuclear recoil yields for S1 and S2.  However, many uncertainties were present, including inaccuracies 
in the modeling of detector threshold effects and the energy dependence of the S1 and S2 yields, known to be non-trivial in liquid xenon 
\cite{NEST_2011} and for which no previously published data in gaseous xenon is known to the authors.

In this study rather than detailing a particular method of extracting information on the nuclear recoil yields, we choose values for the
nuclear recoil yields (informed by the results obtained in \cite{Renner_thesis}), assume a constant energy dependence, and then show that these
assumptions are reasonable by comparison of experimental and Monte Carlo spectra.  For this comparison we use data acquired
with a $^{238}$Pu/Be neutron source (positioned as in experimental setup 2 described in Sec.\ \ref{ss_exp_srccfg}) and a corresponding full Geant4 
\cite{GEANT4} Monte Carlo simulation.  The electronic recoil yields were chosen by demanding consistency with
the results obtained in Sec.\ \ref{s_exp}.  $W_{i} = 24.7$ eV was assumed along with EL gain $g = 734$, and the PMT quantum efficiency $Q$ was adjusted
so that the 662 keV S2 peak in simulation was consistent with that found in experimental data, yielding $Q \approx 17\%$.  
$W_{\mathrm{sc}}$ was adjusted until the intercept $S_{1,0}$ matched that of Fig.\ \ref{fig_exp_s1vstime}, yielding $W_{\mathrm{sc}} = 45.69$ eV.  
The difference between this value of $W_{\mathrm{sc}}$ and that calculated in Sec.\ \ref{s_exp} was used to assign an additional 
systematic uncertainty of 15 eV (see Eq.\ \ref{eqn_exp_Wsc}).
For all recoil energies, the nuclear recoil yields are set to be equal to the electronic recoil yields multiplied
by a quenching factor $\alpha$, where we have chosen for S1 $\alpha_{\mathrm{S1}} = 0.53$ and for S2 $\alpha_{\mathrm{S2}} = 0.17$, corresponding
to nuclear recoil yields of $Y_{1} = \alpha_{\mathrm{S1}}/W_{\mathrm{sc}} = 11.6$ ph/keV for primary scintillation and
$Y_{2} = \alpha_{\mathrm{S2}}/W_{\mathrm{i}} = 6.9$ e$^{-}$/keV for ionization.

The Monte Carlo included the pressure vessel, teflon walls of the field cage and teflon reflector, PMT array, 
mesh grids, and lead block.  A detailed model for the wavelength shifting introduced by the TPB was not implemented, however the teflon
reflectivities on the walls and the teflon reflector were selected such that the z-dependence of S1 matched that of Fig.\ \ref{fig_exp_s1vstime}
in a simulation using axially-incident 662 keV gamma rays, similar to the experimental setup using the $^{137}$Cs source described in Sec.\ \ref{s_exp}.  
The walls were taken to be 100\% reflective and the back reflector to be 79\% reflective.  Though these values are not consistent 
with the expected teflon reflectivity in xenon gas (from \cite{Silva_2010} expected to be 50-60\%), they properly reproduced the linear geometric 
dependence of S1.  The source was modeled by emitting neutrons from a single point behind the lead block in a random direction, 
accompanied by a 4.4 MeV gamma ray also emitted in a random direction.  One neutron and one gamma ray were emitted per event, and the spectrum of 
emitted neutron energies was taken to be that calculated for neutrons produced with a carbon nucleus in the first excited state $^{12}$C$^{*}$ 
(similar to that shown in Fig.\ \ref{fig_neu_alnspectrum}).  The neutron interactions were modeled using the Geant4 high-precision (HP) neutron models, which use 
neutron cross section data from evaluated nuclear data libraries.  The consistency of the Monte Carlo and calculated spectra 
of nuclear recoils can be seen by examining Fig.\ \ref{fig_src_nrspectrum}.

In the simulation, each photon produced via primary scintillation was tracked throughout its entire trajectory, and ionization electrons were
produced in groups called ionization clusters.  The final location of each ionization cluster after drift was calculated, taking into account diffusion,
and the electroluminescent process was modeled by producing a number of photons equal to the EL gain $g$.  Each EL photon was not tracked 
individually but its detection probability was looked up in a table indexed by its $(x,y)$ production location in the 2D EL plane.  This table was 
produced in an independent Monte Carlo run in which 10$^{6}$ photons were generated per point on a grid dividing the EL plane and the detection 
probability for each PMT recorded based on the number of photons collected out of the $10^{6}$ generated.  From the record of photons detected at 
each time in the event, realistic waveforms were constructed for each PMT matching the noise characteristics of the experimental waveforms and 
adding exponential pulses for each photoelectron detected.  The resulting waveforms were passed through a nearly identical analysis to that of 
the experimental data to give results that could be compared directly to experiment.  To simulate the experimental gamma-ray tagging procedure, 
only neutron-induced events, those in which a neutron scattered inelastically on any material in the simulation or elastically on xenon, were 
considered in the analysis.  To roughly match the detector threshold effects observed at low S1 values in experiment, a peak-finding threshold 
was chosen appropriately in the Monte Carlo analysis.

\begin{figure*}[!htb]
\includegraphics[scale=0.55]{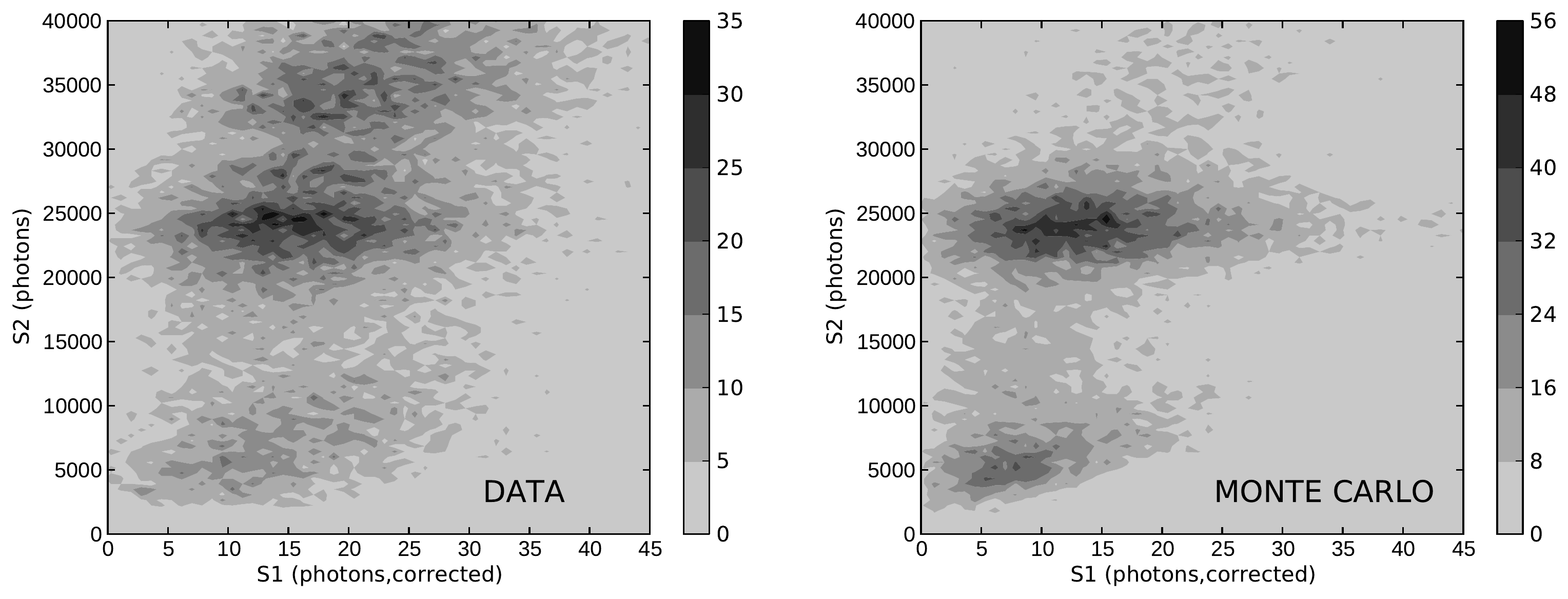}
\caption{\label{fig_neu_dvsmcs2s1}S2 vs. S1 for low-energy events in experiment and Monte Carlo simulations generated based on the selected constant nuclear 
recoil yields.}
\end{figure*}

Figure \ref{fig_neu_dvsmcs2s1} shows the S2 signals plotted against S1 corrected for z-dependence in both data and Monte Carlo.  Smearing was applied to
the S1 and S2 values determined in Monte Carlo to account for geometric dependencies in the detection that degraded the experimental energy resolution,
which appeared to influence more heavily S1.  The analysis included tagging, single-pulse, diffusion, and radial cuts.  Similar cuts were 
made in the corresponding Monte Carlo run, and the good
qualitative agreement, particularly in the location of the nuclear recoil band, shows that the nuclear recoil quenching factors
employed in the simulation are reasonable.  One significant discrepancy in the Monte Carlo results is the relatively 
fewer number of events in the band lying at approximately 35000 S2 photons which is due
to 40 keV gamma rays produced by neutron inelastic scattering on $^{129}$Xe.  This is likely due to the fact that only neutrons accompanied by a 4.4 MeV gamma
(energies 2-6 MeV) are considered in the Monte Carlo, while in the experimental data some fraction of the neutrons accompanied by the 7.7 MeV gamma 
(energies $< 3$ MeV) are also included.  This is because the 7.7 MeV gamma may not fully deposit its energy in the NaI scintillator and therefore 
may give an energy in the window of allowed NaI energies (2.4-5.0 MeV for the experimental run described in this section, and similar for that described
in Sec.\ \ref{ss_neu_analysis} - \ref{ss_neu_yields}).   The inelastic $\mathrm{n} \, + \, ^{129}$Xe 
scattering cross section is significantly higher \cite{ENDF} for these lower energy neutrons, so one 
should expect relatively more 40 keV gamma rays in the experimental data.  However, the nuclear recoils produced by these lower energy neutrons will be lower 
in energy and therefore should not contribute significantly to the nuclear recoil spectrum at the energies observable in the data.
\section{Conclusions\label{s_ccl}}
We have simultaneously observed ionization and scintillation produced by nuclear recoils in gaseous xenon.
It was confirmed that xenon in the gas phase, as in the liquid phase, is capable of distinguishing nuclear and electronic recoils based on the ratio of
observed scintillation to ionization.  It appears that relative to the corresponding yields for electronic recoils, the primary scintillation yield for nuclear recoils is
quenched by a factor of roughly 2, and the ionization yield by a factor of roughly 5.  Further investigation is required to determine precise nuclear recoil 
yields and fully investigate the potential advantages offered by gas phase operation.  The greatest impact of the present results is the ability to inform 
future measurements of nuclear recoils in gaseous xenon.  

The presence of gamma depositions in regions from which S2 cannot be collected provides an evident background and is known to also be of concern in the 
liquid phase.  This is likely to be of even greater concern in the gas phase, as gamma rays can travel farther, and is evident in the inability to
eliminate events consisting only of a single fluorescent x-ray.  In such events, x-rays produced outside of the active region were able to travel into the
active region and interact there.  Such events can in principle be eliminated with a fiducial cut, but this would require a larger detector than the one
used in the present experiment.

Though only estimates of the nuclear recoil yields were given, they can be used to predict the necessary light
collection efficiency required to observe recoils of a given energy of interest.  With 3\% light collection efficiency at the EL plane, we were not able to see many recoils
with energy less than approximately 30 keV.  Therefore a $\sim$10\% light collection efficiency at least would be necessary to perform a stronger measurement of the nuclear
recoil yields down to near 10 keV.  The higher photon statistics will also help in characterizing the electronic/nuclear recoil discrimination power based
on S2/S1 that is possible in the gas phase.  The TPB was found to be a necessity for achieving enough light collection efficiency to identify the 
nuclear recoils, and in the future more sophisticated ideas such as the use of light guides coupled to PMTs \cite{Nygren_2013} may be necessary to achieve the light
collection regime of interest.  

\appendix*
\section{Radioisotope Neutron Sources\label{s_src}}
\subsection{Neutron Production Mechanism}
The neutron sources used in this study all consist of an $\alpha$-emitting radioactive isotope mixed with beryllium ($^{9}$Be) and generate 
neutrons based on the $(\alpha,\mathrm{n})$ reaction \cite{Knoll_2000}

\begin{equation}\label{eqn_src_alphan}
 \alpha + ^{9}\mathrm{Be} \rightarrow ^{12}\mathrm{C} + \mathrm{n}.
\end{equation}

The Q-value of the reaction is Q$ = 5.701$ MeV, and this energy is released in the form of neutron kinetic energy, neglecting any carbon recoil
kinetic energy, unless the carbon nucleus is left in an excited state, in which case some of the energy is emitted in the form of a gamma ray
in coincidence with the neutron \footnote{The mean lifetimes of the excited carbon nuclear states of interest here are so short ($< 0.1$ ps) 
compared to the measurable time scale that the gamma can be considered to be emitted simultaneously with the neutron for practical purposes.}.
If left in the first excited state $^{12}$C$^{*}$, a gamma ray of energy 4.439 MeV is emitted, and if left in the second excited state $^{12}$C$^{**}$, 
a gamma ray of energy 7.654 MeV is emitted \cite{ENSDF}.  In this study we detect the coincident gamma and include it in the acquisition trigger to tag 
neutron-emitting decays and thereby significantly reduce the number of background events acquired.  

We describe here how to calculate the neutron spectra of radioisotope sources under the assumption that their active regions consist of a uniform volume of 
$^{9}$Be throughout which the $\alpha$-emitting isotope is uniformly distributed.  We also assume that 
the total number of $\alpha$-emitting isotope atoms present is much less than the total number of beryllium atoms in the mixture, so that each emitted 
$\alpha$ can be considered to interact with only atoms of beryllium.  In addition, we do not consider neutrons produced due to the break-up
reaction $\alpha + ^{9}$Be $\rightarrow \alpha' + ^{9}$Be$^{*}$, $^{9}$Be$^{*} \rightarrow ^{8}$Be $+$ n \cite{Geiger_1975}.  These neutrons lie at lower 
energies $\lesssim 3$ MeV, and they will not be observed in the adopted trigger scheme as no coincident gamma ray is emitted.  Our calculations follow those 
of \cite{Geiger_1975}, \cite{Vijaya_1973}, and \cite{Kumar_1977}, and make use of $(\alpha,\mathrm{n})$ cross sections from the Japanese Evaluated
Nuclear Data Library (JENDL) \cite{JENDL} and neutron-xenon scattering cross sections from the Evaluated Nuclear Data File (ENDF) \cite{ENDF}.  The cross
sections were processed using the tools developed in \cite{Mattoon_2012}.  Table
\ref{tbl_src_isotopes} gives information on the decays of isotopes $^{238}$Pu and $^{241}$Am used as $\alpha$-emitters in the sources used in this study.

\begin{table}
\caption{\label{tbl_src_isotopes}Selected radioactive decay products of $^{238}$Pu and $^{241}$Am.  All data shown is from \cite{LBLTOI}.  Both isotopes 
emit alpha particles with 
an average energy of approximately 5.5 MeV.  Gamma ray emission from $^{238}$Pu is negligible, while some low-energy gamma rays are emitted by $^{241}$Am.  In 
particular, the 60 keV gamma ray is likely to produce significant background without sufficient shielding of the source.}
\begin{ruledtabular}
\begin{tabular}{lccc}
Isotope, $\tau_{1/2}$, Q & Product & Energy (keV) & \% BR\\
\hline
$^{238}$Pu & $\alpha$ & 5357.7 & 0.105\\
$\tau_{1/2} = 87.7 \pm 0.3$ yr & $\alpha$ & 5456.3 & 28.98\\
$\mathrm{Q} = 5593.20 \pm 0.19$ & $\alpha$ & 5499.03 & 70.91\\
\hline
$^{241}$Am & $\gamma$ & 13.946 & 9.6\\
$\tau_{1/2} = 432.2 \pm 0.7$ yr & $\gamma$ & 59.5412 & 35.9\\
$\mathrm{Q} = 5631.81 \pm 0.12$ & $\alpha$ & 5388.23 & 1.6\\
& $\alpha$ & 5442.80 & 13.0\\
& $\alpha$ & 5485.56 & 84.5\\
\end{tabular}
\end{ruledtabular}
\end{table}

\subsection{Spectrum of Emitted Neutrons}\label{ss_nspectrum}
The neutron production process proceeds as follows.  The $\alpha$-emitting isotope decays, yielding an $\alpha$ particle of mass $m_{\alpha}$ 
that travels through the surrounding medium of beryllium atoms of mass $m_{b}$, losing energy and often stopping completely without undergoing the reaction 
shown in Eq.\ \ref{eqn_src_alphan}.  However, some alpha particles (of order 1 in $10^{4}$) will undergo the neutron-producing reaction of interest at 
an energy of $E_{\alpha}$ and generate a neutron of mass $m_{n}$ at some angle $\theta$ in the center of mass frame of the interaction that can be directly 
related to the emitted neutron energy $E_{n}$ via \cite{Vijaya_1973}

\begin{equation}\label{eqn_src_costheta}
 \begin{split}
  \cos\theta = & E_{n}(m_{n}+m_{c})^{2} - E_{\alpha}(m_{b}m_{c} + m_{n}m_{\alpha})\\ 
  & \frac{\,\,\,\,\,\,\,\,\,\,\,\,\,\,\,\,\,\,\,\,\,\,\,\,\,\,\,\,\,\,\,\,\,\,\,\,\,\,\,\,\,\,\,\,\,\,\,\,\,\,\,\,\,\,\,\,\,\,\,\, 
  - Q'm_{c}(m_{n}+m_{c})}{2[E_{\alpha}m_{\alpha}m_{n}(E_{\alpha}m_{b}m_{c} + Q'm_{c}(m_{n}+m_{c}))]^{1/2}},
 \end{split}
\end{equation}

\noindent where $Q'$ in this case is the energy released in the reaction in the form of neutron kinetic energy and may be equal to the full $Q$-value
of 5.701 MeV or the full Q-value minus the energy of excitation left with the resulting carbon nucleus that is emitted in the form of a gamma ray.
The angular distribution of neutrons for a given alpha energy $E_{\alpha}$ can thus be expressed in terms of the neutron energy $E_{n}$, and the number of
neutrons emitted with energy in an interval $(E_{n},E_{n}+dE_{n})$ can be written as

\begin{equation}\label{eqn_src_probEn}
 \begin{split}
  G &(E_{n};E_{\alpha}) dE_{n} = \\ & \frac{1}{\sigma_{T}(E_{\alpha})}\frac{d\sigma(E_{n};E_{\alpha})}{d\Omega}\cdot\frac{4\pi}{E_{n}(\theta = 0) - E_{n}(\theta = \pi)} \, dE_{n},
 \end{split}
\end{equation}

\begin{figure}
\includegraphics[scale=0.48]{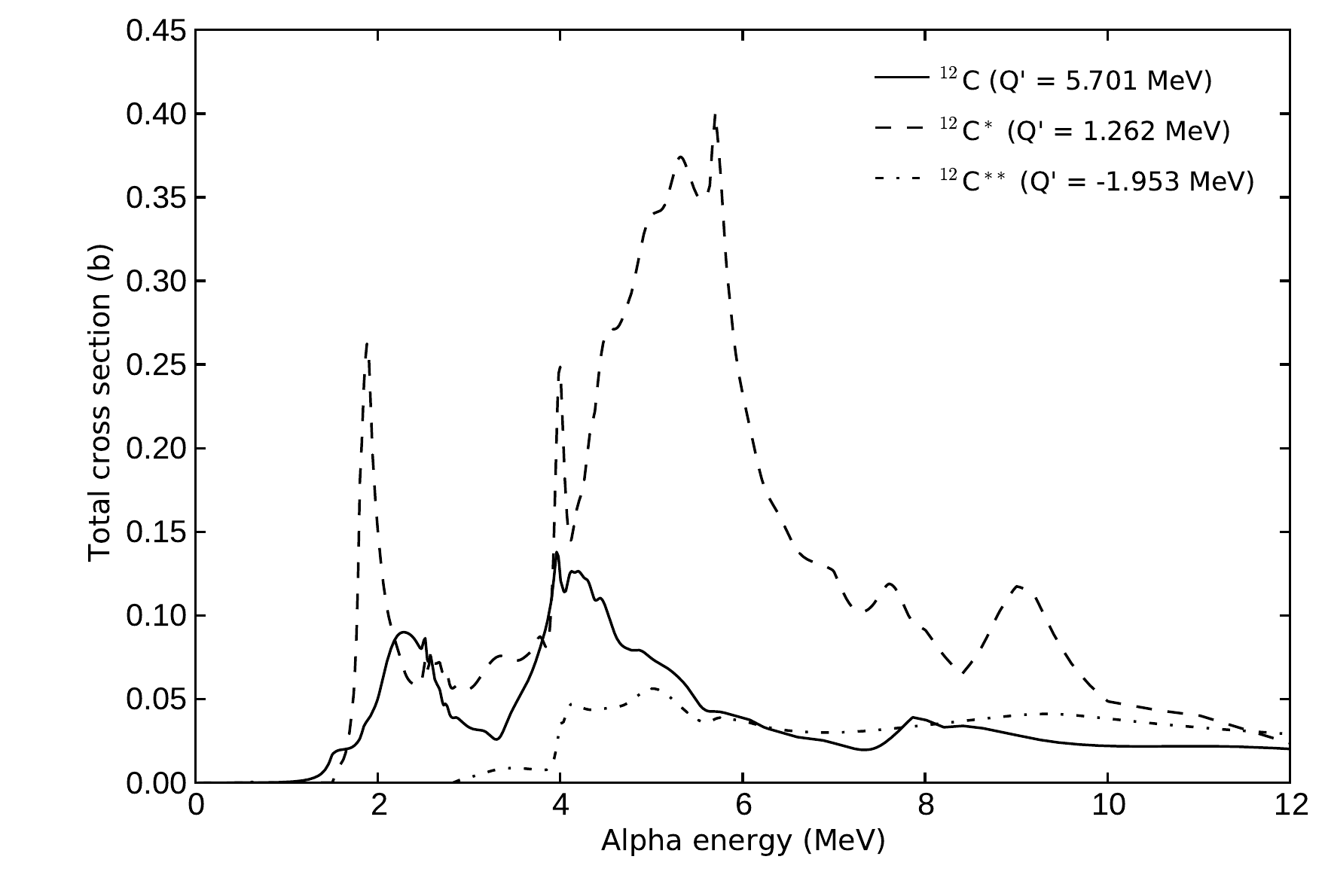}
\caption{\label{fig_src_sigmaT}Total cross section for the reaction $^{9}$Be$(\alpha,\mathrm{n})^{12}$C.  All data shown in this figure are from
JENDL \cite{JENDL}.}
\end{figure}

\noindent where $\sigma_{T}(E_{\alpha})$ is the total cross section of the $(\alpha,\mathrm{n})$ reaction for alpha energy $E_{\alpha}$, shown in 
Fig.\ \ref{fig_src_sigmaT}, and $d\sigma/d\Omega$ is the differential cross section.  The number of neutrons with energy $E_{n}$
produced for emitted alpha $i$ of energy $E_{\alpha,i}$ can then be determined by multiplying $G(E_{n};E_{\alpha})$ by the probability 
of interaction at alpha energy $E_{\alpha}$ along the track, that is, 
$\rho\sigma_{T}(E_{\alpha})dx = \rho\sigma_{T}(E_{\alpha})(dx/dE_{\alpha})dE_{\alpha}$, and integrating over all energies from $0$ to $E_{\alpha,i}$,

\begin{equation}\label{eqn_src_neutronspec}
 \begin{split}
  N_{i} &(E_{n}) = \\ & \int_{0}^{E_{\alpha,i}} \frac{4\pi [d\sigma(E_{n};E_{\alpha})/d\Omega]}{dE_{\alpha}/(\rho dx)[E_{n}(\theta = 0) - E_{n}(\theta = \pi)]} dE_{\alpha}.
 \end{split}
\end{equation}

The final spectrum will be a sum of such integrals over the emitted $\alpha$ energies of the source weighted by their branching ratios $x_{i}$,

\begin{equation}\label{eqn_src_neutronspec_sum}
 N(E_{n}) = \sum_{i}x_{i}N_{i}(E_{n}).
\end{equation}

Figure \ref{fig_neu_alnspectrum} of Sec.\ \ref{ss_nrspec} shows the spectrum for a $^{241}$Am/Be source calculated assuming the alpha particle energies and branching ratios from 
Table \ref{tbl_src_isotopes}.  Note that this calculation is only valid for $dx \ll \lambda_{\alpha,\mathrm{n}}$, where $\lambda_{\alpha,\mathrm{n}}$ 
is the mean interaction length for the $(\alpha,\mathrm{n})$.  However, the average length of the track produced by a 5.5 MeV alpha particle in beryllium (density 
$\rho = 1.23\, \times 10^{23}$ cm$^{-3}$ \cite{RPP_2012}) can be calculated using the alpha stopping power from \cite{NIST_ASTAR} as 
$\int (dx/dE) dE \approx 28 \mu$m.  Using total cross section $\sigma_{T} < 0.4$ barns (see Fig.\ \ref{fig_src_sigmaT}), 
$\lambda_{\alpha,\mathrm{n}} = (\rho\sigma_{T})^{-1} > 20$ cm, and therefore this condition holds.

\subsection{Resulting Spectrum of Nuclear Recoils}

\begin{figure}
\includegraphics[scale=0.48]{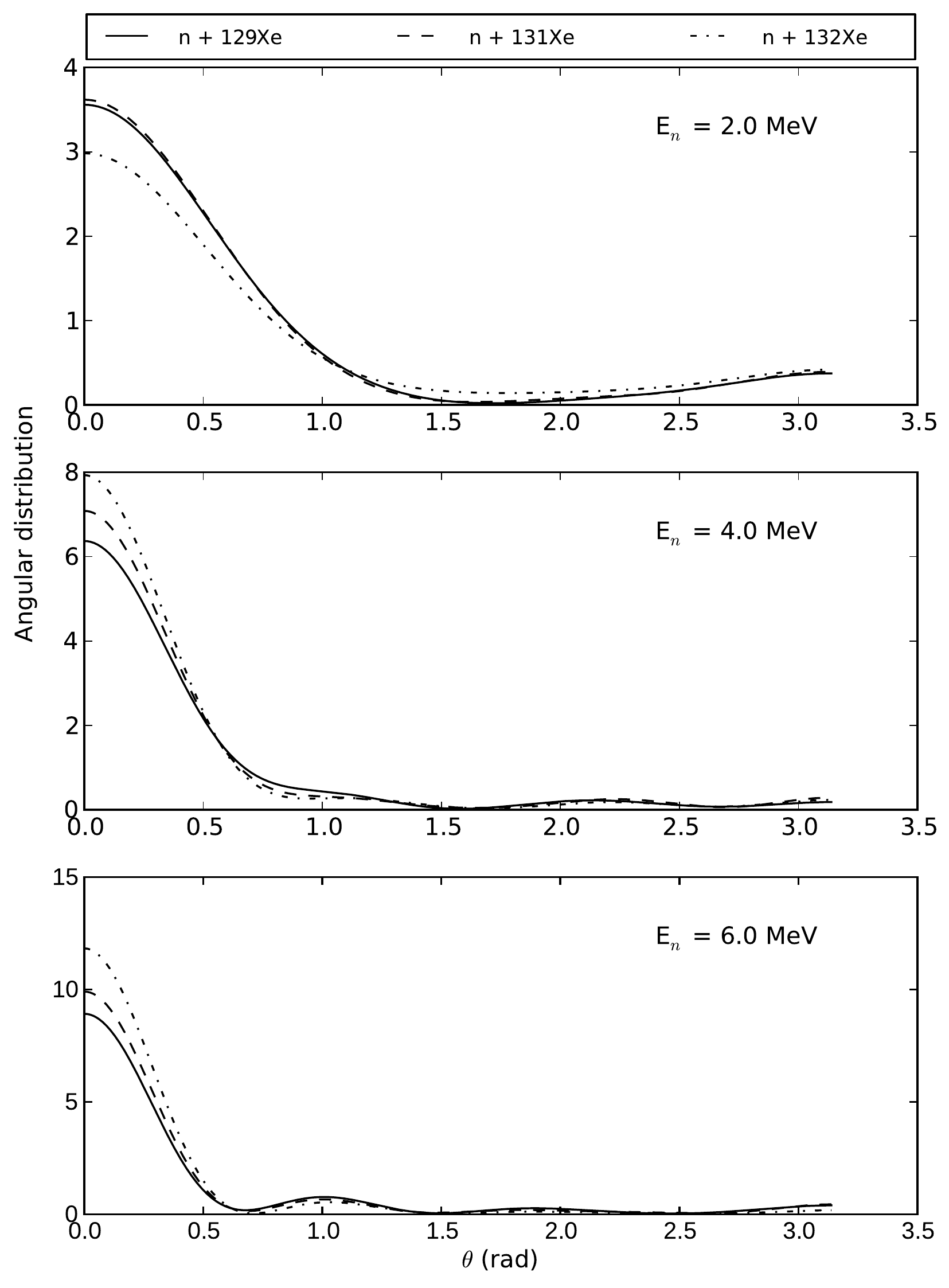}
\caption{\label{fig_src_nxedcs}Angular distributions (data from \cite{ENDF}) for elastic neutron scattering, n $+$ Xe $\rightarrow$ n$'$ $+$ Xe$'$, 
for three different neutron energies incident on three different xenon isotopes.  Note the strong peak in the forward direction, followed by one or several peaks at higher 
angles which are responsible for the structure in the nuclear recoil spectrum at higher energies.}
\end{figure}

\begin{figure}
\includegraphics[scale=0.48]{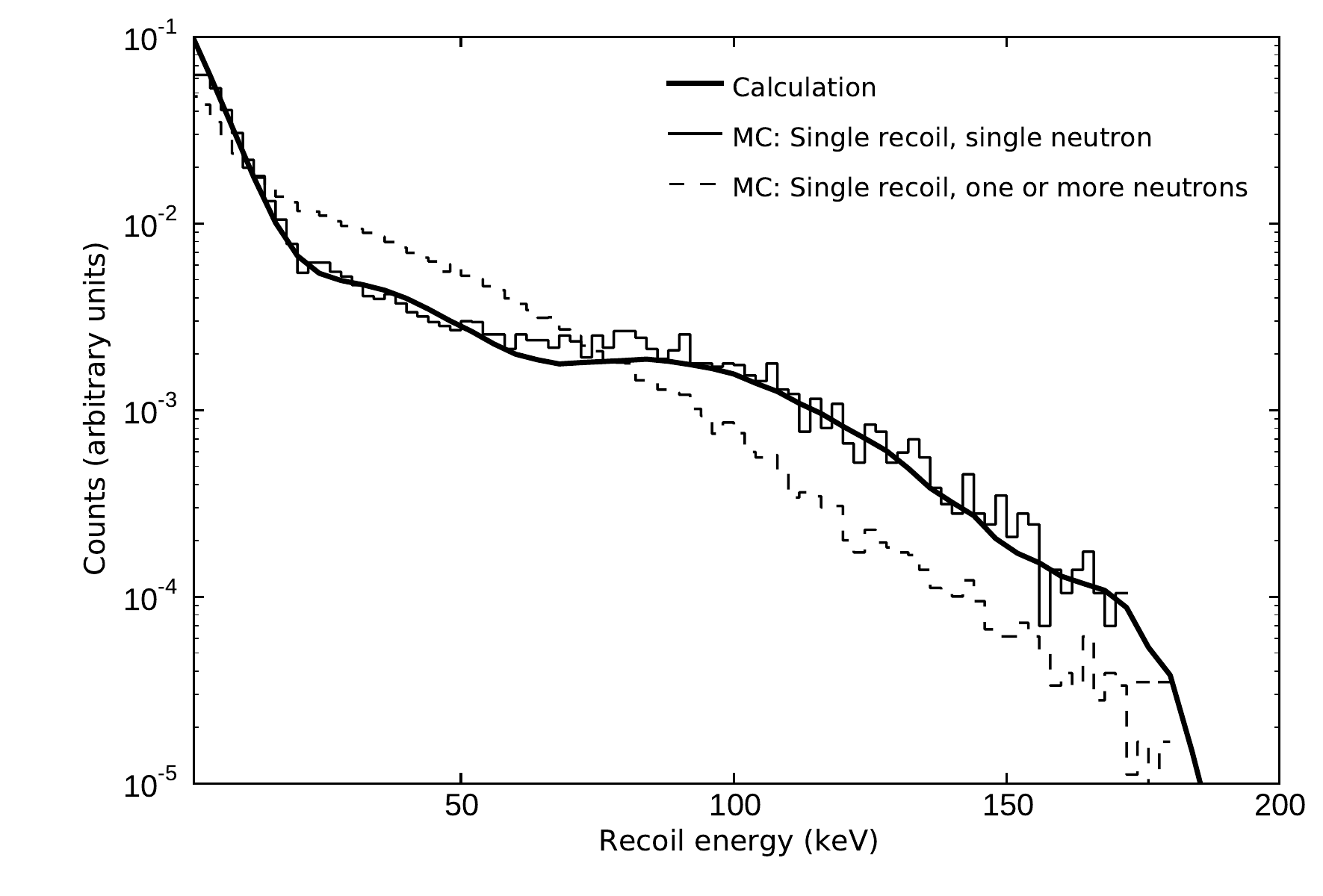}
\caption{\label{fig_src_nrspectrum}Spectrum of nuclear recoils assuming neutrons emitted from a $^{238}$Pu/Be source in which the carbon
nucleus resulting from the $^{9}$Be$(\alpha,\mathrm{n})^{12}$C reaction was left in the first excited state.  The spectrum was obtained from a
calculation using Eq.\ \ref{eqn_src_nrspectrum} and Eq.\ \ref{eqn_src_nrspectrum_sum} (thick solid line) and from a Geant4 Monte Carlo.  Two curves are
shown from the same Monte Carlo run.  The solid line is the spectrum considering only events containing a single xenon nuclear recoil and a single neutron.  The 
dashed line is the spectrum of events containing a single xenon nuclear recoil, but in which the event may have contained multiple neutrons produced by interaction
of the incident neutron with the lead.  The presence of these secondary neutrons significantly distorts the spectrum of xenon nuclear recoils.}
\end{figure}

From a spectrum of emitted neutrons, one can construct a spectrum of nuclear recoils produced when those neutrons are incident on a volume
of xenon atoms by knowing the cross section for neutron elastic scattering (differential with respect to solid angle).  Similar to 
Eq.\ \ref{eqn_src_costheta}, we can express the energy of a nuclear recoil in terms of the cosine of the
scattering angle in the center-of-mass frame of the neutron-nucleus collision as

\begin{equation}\label{eqn_src_erecoil}
 E_{x'} = \frac{2E_{n}m_{n}m_{x,a}}{(m_{n}+m_{x,a})^{2}}(1 - \cos\theta),
\end{equation}

\noindent where $m_{x,a}$ is the mass of the target xenon nucleus and $a$ is an index corresponding to the isotope of the nucleus.

Similar to Eq.\ \ref{eqn_src_probEn}, we can use this relation to write the angular distribution of scattered neutrons in terms of the recoil 
energy, and determine the number of neutron interactions per unit recoil energy $N_{r,a}(E_{x'};E_{n})$ yielding a nuclear recoil with energy in the interval 
$(E_{x'},E_{x'}+dE_{x'})$ for a neutron of energy $E_{n}$ incident on a xenon nucleus of isotope $a$,

\begin{equation}\label{eqn_src_nrspectrum}
 \begin{split}
  N_{r,a} &(E_{x'}; E_{n}) =\\
  & \frac{(m_{n}+m_{x,a})^{2}}{m_{n}m_{x,a}}\cdot \frac{\pi}{E_{n}} \Biggl[\frac{d\sigma(E_{x'};E_{n})}{d\Omega}\Biggr]_{a} \rho\Delta x,
 \end{split}
\end{equation}

\noindent where $\rho$ is the density of the xenon volume and $\Delta x$ is the thickness of xenon traversed by the neutron.  
Figure \ref{fig_src_nxedcs} shows the angular distribution for neutrons scattered off of several different xenon isotopes.  Note that because we will
be interested in the normalized spectrum, and $\rho\Delta x$ is a constant independent of the isotope, it can be absorbed into an overall normalization
factor in the final spectrum, which is a weighted superposition of individual nuclear recoil spectra summed over xenon isotopes and integrated over neutron 
energies,

\begin{equation}\label{eqn_src_nrspectrum_sum}
 N(E_{x'}) = \sum_{a}f_{a}\int_{0}^{\infty}N(E_{n})N_{r,a}(E_{x'}; E_{n})dE_{n}.
\end{equation}

\noindent where the $f_{a}$ are the fractional compositions of natural xenon for each isotope and $N(E_{n})$ is the spectrum of emitted neutrons from 
Eq. \ref{eqn_src_neutronspec_sum}.

Figure \ref{fig_src_nrspectrum} shows the calculated spectrum of nuclear recoils assuming the neutron spectrum is that emitted by the source with a
coincident 4.4 MeV gamma ray, corresponding to the dashed curve in Fig.\ \ref{fig_neu_alnspectrum}.  A target of natural xenon is assumed with fractional 
isotopic composition taken from the ``representative'' values reported in \cite{Bievre_1993}.  Isotopes $^{124}$Xe and $^{126}$Xe are excluded from the
calculation due to lack of cross section data but have negligible natural abundances.  The spectrum is compared with results from a Geant4 Monte Carlo 
of the experimental setup in which a source emitting the same neutron spectrum input to the calculation was positioned behind a 2-in.\ thick lead block.  The calculated
and Monte Carlo recoil spectra agree well when considering only events in which a single xenon recoil occurred and only one neutron was present throughout 
the entire event.  Including also events in which the emitted neutron interacted in the lead to produce additional neutrons 
gives a significantly altered spectrum.

\begin{acknowledgments}
This work was supported by the following agencies and institutions: the Director, Office of Science, Office of Basic Energy Sciences, of the U.S. Department 
of Energy, and the National Energy Research Scientific Computing Center (NERSC), supported by the Office of Science of the U.S. Department of
Energy, both under Contract No.\ DE-AC02-05CH11231; the European Research Council under the Advanced Grant 339787-NEXT; the Ministerio de Econom\'{i}a y 
Competitividad of Spain under Grants CONSOLIDER-Ingenio 2010 
CSD2008-0037 (CUP), FPA2009-13697-C04-04, FPA2009-13697-C04-01, FIS2012-37947-C04-01, FIS2012-37947-C04-02, FIS2012-37947-C04-03, and FIS2012-37947-C04-04; 
and the Portuguese FCT and FEDER through the program COMPETE, Projects 
PTDC/FIS/103860/2008 and PTDC/FIS/112272/2009.  J. Renner acknowledges the support of a Department of Energy National Nuclear Security Administration 
Stewardship Science Graduate Fellowship, grant number DE-FC52-08NA28752.
\end{acknowledgments}

\bibliography{bib/xenon_neutrons}

\end{document}